\documentclass[
    onecolumn,
    preprintnumbers,
    amsmath,
    amssymb,
    prd,
    showpacs
]{revtex4}
\usepackage{graphicx}%
\usepackage{amsmath}
\usepackage[normalem,normalbf]{ulem}
\usepackage{amssymb}
\usepackage{bm}
\usepackage{enumerate}
\newcommand{\nn}{\nonumber}

\newcommand{\cA}{{\hat {\cal A}}}
\newcommand{\cH}{{\hat {\cal H}}}
\newcommand{\ca}{\hat{\mathfrak{a}}}
\newcommand{\hA}{\hat A_1}
\newcommand{\dsp}{\displaystyle}
\newcommand{\cE}{{\cal E}}
\newcommand{\sign}{\mbox{sign}}
\begin{document}
\title{A new variational perturbation method for double well oscillators}
\author{Hyeong-Chan Kim}
\email{hckim@phya.yonsei.ac.kr}
\author{Jae Hyung Yee}%
\email{jhyee@phya.yonsei.ac.kr}
\affiliation{Dept. of Physics, Yonsei University, Seoul Republic
of Korea.
}%
\date{\today}%
\bigskip
\begin{abstract}
\bigskip

We propose a variational perturbation method based on the observation
that eigenvalues of each parity sector of both the anharmonic and
double-well oscillators are approximately equi-distanced. The
generalized deformed algebra satisfied by the invariant operators of the
systems provides well defined Hilbert spaces to both of the oscillators.
There appears a natural expansion parameter defined by the ratios of
three distance scales of the trial wavefunctions. The energies of the
ground state and the first order excited state, in the $0^{th}$ order
variational approximation, are obtained with errors $<10^{-2}$\% for
vast range of the coupling strength for both oscillators. An iterative
formula is presented which perturbatively generates higher order
corrections from the lower order invariant operators and the first order
correction is explicitly given.

\end{abstract}
\maketitle

\section{Introduction}
There are two well known problems in studying the
anharmonic and the double-well oscillators.
The first is that a naive perturbation theory of anharmonic oscillator
leads to a divergent perturbation series even for an infinitesimal value
of the coupling strength due to the eventual dominance of the
perturbative correction over the un-perturbed contribution for large
amplitude of oscillation~\cite{bender}.
To remedy this difficulty several generalized perturbation methods have
been developed~\cite{friedberg,justin}.
There have been many recent developments which provide well defined
perturbative iterative solutions for the eigenvalues and
eigenfunctions~\cite{mei,weinstein,friedberg,mahapatra,turbiner}.
An interesting differential equation method, in parallel with Mathieu
equation, was presented by Liang and M\"{u}ller-Kirsten~\cite{liang}.
A through analysis for this asymptotic perturbation and large order
behaviors was given in Ref.~\cite{muller}.

Another difficulty is that a double-well oscillator has completely
different algebraic structure and ground state wavefunction from those
of anharmonic oscillator.
Therefore, most perturbative approaches treat the two oscillators
separately~\cite{weinstein,friedberg}.
The transition from a single-well system to a double-well system of a
time-dependent oscillator was calculated through a nonperturbative
method such as the WKB approximation~\cite{bhattacharya,gildener} or by
accommodating contributions from many excited states in addition to the
ground states of symmetric oscillator~\cite{banerjee}.
Even though these methods provide good approximations to the ground
state and excited state energies, its generalization to field theory of
infinite degrees of freedom or to time-dependent systems with
symmetry-breaking-like phenomena encounters difficulties.
Therefore, it is important to establish a consistent approximation
method that can treat both of the oscillators in the same way.
This subject has not been extensively studied compared to the problem of
divergent perturbation.

The purpose of the present paper is to present a possible
resolution to the second problem using a variational
perturbation method.
We present a unified treatment of the anharmonic and the double-well
oscillators in the point of view of a generalized deformed oscillator.
This provides an algebraic method in which the two oscillators are dealt
with on an equal footing.
We also present a natural expansion parameter defined by the ratios of
three length scales appearing in the trial wavefunction.
Its numerical value is smaller than $1/2$  for all physical parameter
range of the system for the double-Gaussian trial wavefunctions used in
this paper.
To do these, we solve the Liouville-von Neumann (LvN) equation
perturbatively.

The generalized deformed oscillator~\cite{bonatsos} was
developed in the approach of quantum group
theory~\cite{drinfeld,chai}.
The algebra is generated by the operators $\{1, a, a^\dagger, N\}$ and
the structure function $\Phi(x)$, satisfying the relations
\begin{eqnarray} \label{gen:osc}
~[a, N]&=&a, ~~[a^\dagger ,N]= -a^\dagger, \\
a^\dagger a&=& \Phi(N)=[N], ~~~ a a^\dagger=
\Phi(N+1)=[N+1],\nn
\end{eqnarray}
where $N$ is the number operator and $\Phi(x)$ is a
positive analytic function with the condition
\begin{eqnarray*} \label{Phi:0}
\Phi(0)=0.
\end{eqnarray*}
This ensures the vacuum state $|0\rangle$ to be defined as the
eigenstate of $N$ with zero eigenvalue and satisfies
\begin{eqnarray*} \label{0:gen}
a|0\rangle =0 .
\end{eqnarray*}
It can be proved that the generalized deformed algebra
provides a Hilbert space of eigenvectors $|0\rangle,
~|1\rangle,~\cdots , |n\rangle,\cdots$ of the number
operator $N$ with $\langle n|m\rangle = \delta_{mn}$.
These eigenvectors are generated by the formula,
\begin{eqnarray*} \label{nptl}
|n\rangle = \frac{1}{\sqrt{[n]!}}\left( \hat a^\dagger
\right)^n
    |0\rangle,~~~
[n]! = \prod_{k=1}^n [k]= \prod_{k=1}^n \Phi(k) .
\end{eqnarray*}
The operators $a^\dagger $ and $a$ are the creation and the
annihilation operators of this deformed oscillator:
\begin{eqnarray*} \label{ada:gen}
\hat a|n\rangle= \sqrt{[n]} |n-1\rangle, ~~~\hat
a^\dagger|n\rangle=\sqrt{[n+1]} |n+1\rangle .
\end{eqnarray*}
The harmonic oscillator is obtained for $\Phi(n)=n$ and the $q-$deformed
oscillator for $\displaystyle\Phi(n)=\frac{q^n-q^{-n}}{q-q^{-1}}$ with
the Hamiltonian, $\displaystyle H=\frac{\hbar \omega}{2}(\hat a\hat
a^\dagger +\hat a^\dagger \hat a)$.

The Hamiltonian describing both the anharmonic oscillator
and the double-well oscillator is given by
\begin{eqnarray} \label{H:ah}
H= \frac{\hat p^2}{2}+ \frac{\omega^2}{2} \hat
x^2+\frac{\lambda}{4} \hat x^4,
\end{eqnarray}
with $\omega^2$ being a positive (anharmonic) or a negative
(double-well) real number.
For $\lambda \rightarrow 0$ with $\omega^2 >0$, the system becomes a
harmonic oscillator (harmonic oscillator limit) and for $\omega^2 <0$
with $\lambda \rightarrow 0_+$ the system describes a set of two
separated wells of harmonic oscillators (infinitely separated
double-well limit).

The LvN equation~\cite{lewis},
\begin{eqnarray} \label{LvN} 0\equiv
i \hbar \frac{d \hat A(\hat p,\hat x,t)}{dt} = i \hbar
\frac{\partial
    \hat A}{\partial t}+[\hat A,H]
\end{eqnarray}
is equivalent to the Schr\"{o}dinger equation and physical
information can be obtained from the invariant operator
$\hat A$.
Based on this operator, a complete Hilbert space of the oscillator can
be constructed.
The equation~(\ref{LvN}) was used as a starting point in developing many
approximation methods and of physical applications~\cite{kim,bak}.
This method provides a powerful basis in constructing the variational
perturbation methods since the zeroth order solution of~(\ref{LvN}) can
be made to be a variational approximation of the system.
In the present paper, we will utilize this aspect to develop a new
variational perturbative approximation with two variational parameters
for time-independent oscillators.
Generalization to time-dependent systems is straightforward since the
LvN equation is known to be one of the best ways to treat the
time-dependent system~\cite{lewis,kim}.

Standard perturbation theory based on harmonic oscillators assumes that
the coupling term $\lambda \hat x^4$ is smaller than the other term:
$\langle \lambda \hat x^4\rangle \ll \langle \omega^2 \hat x^2\rangle$.
This leads to a `localization condition' of the expectation value,
$\displaystyle \langle\hat x^2\rangle\simeq \frac{\hbar}{2\omega} \ll
\omega^2/(3\lambda)$, with respect to the ground state wavefunction.
This inequality restricts the applicability of the standard perturbative
method to small coupling.
To overcome this limitation, the Gaussian approximation
method~\cite{sypi} and its generalizations have been
developed~\cite{oko,hckim2}.
The Gaussian approximation starts from defining a new frequency scale
$\Omega_G$, which is variationally determined to satisfy
\begin{eqnarray*}
 \Omega_G^2
=\omega^2+\frac{3\lambda}{2\Omega_G}.
\end{eqnarray*}
The localization condition now takes the form
$\displaystyle \frac{3\lambda}{2\Omega_G^3} \ll 1$, which
is satisfied for $\omega^2>0$.
The characteristic length scale and momentum scale corresponding to this
frequency are
\begin{eqnarray*}
x_G=\sqrt{\frac{\hbar}{2\Omega_G}},~~ p_G= \sqrt{\hbar
\Omega_G \over 2}.
\end{eqnarray*}
The Gaussian wave-function is annihilated by the
annihilation operator,
\begin{eqnarray*}
\hat a_G=\frac{i\hat p}{2p_G }+\frac{\hat x}{2x_G},
\end{eqnarray*}
which satisfies a simple commutation relation $[\hat a_G,
\hat a_G^\dagger]=1$.
Based on this commutation relation, a complete Hilbert space can be
constructed by consecutive operations of the creation operator $\hat
a_G^\dagger$ to the ground state $|0\rangle_G$ defined by $\hat
a_G|0\rangle_G=0$.

There have also been some attempts to apply the Gaussian
approximation to the double-well systems, in which one
wants to find an annihilation operator of the form
\begin{eqnarray*}
\hat a_{x_0}= \frac{i \hat p}{2p_G}+ \frac{\hat x-x_0}{2x_G},
\end{eqnarray*}
where $x_0$ and $\Omega_G$ are the parameters to be
determined by variation~\cite{mahapatra,weinstein}.
In spite of the success of this approach in finding an approximate
ground state energy of double-well oscillator, it is to be noted that
the approach does not reflect the symmetry of the system in the sense
that the state annihilated by $\hat a_{x_0}$ does not have the symmetry
of the potential, the space inversion symmetry.
An approximation method which keeps this symmetry is the Double Gaussian
Approximation (DGA)~\cite{cooper,kovner,kim2}, in which the sum of two
Gaussian wave-packets (or similar ones) are used to approximate the
ground state wave-function of a double-well oscillator.
A major deficiency of the DGA is that no Hilbert space based on the
approximate ground state has been found, which is one of the issues to
be dealt with in the present paper.

We develop a method in which a double-well oscillator is dealt with in
the same way as a single-well anharmonic oscillator.
Before introducing the method, we point out some interesting features of
double-well and anharmonic oscillator systems.
The first is that the eigenvalues of a double-well oscillators are not
approximately equi-distanced contrary to the case of an anharmonic
oscillator.
For example, the difference between the first excited state energy and
the ground state energy is considerably different from the difference
between the second excited state energy and the first excited state
energy.
This implies that a single set of creation and annihilation operators
$(\hat a,~\hat a^\dagger)$ cannot connect all the eigenstates with
reasonable accuracy.
Considering the parity even states and odd states separately, however,
one finds that the energy eigenvalues are almost even-spaced in each
parity sector [For example, see chapter 5 of Ref.~\cite{merz} for exact
solution of double-well oscillator with the potential
$\frac{1}{2}(k|x|-a)^2$].
This suggests us to consider operators connecting eigenstates by two
steps, connecting the states among the same parity eigenstates.
We call the ground state $|0\rangle$ and the $1^{st}$ excited state
$|1\rangle$ as the `ground' states for each parity sector.
Thus, the $1^{st}$ excited state $(|0_O\rangle\equiv |1\rangle)$ is the
ground state of odd parity sector and the state $|0_E\rangle=|0\rangle$
is the ground state of even parity sector.

Another interesting feature of double-well oscillator is that there
exists a set of operators ($\hat A_0,~\hat A_0^\dagger$) and even and
odd ground states $(|0_{E/O}\rangle)$, which consistently describe both
the harmonic oscillator limit and the infinitely separated double-well
limit.
%
%
Consider the following states of even and odd parities,
\begin{eqnarray*} \label{vac}
|0_E\rangle_0= \frac{1}{N_E}\frac{|+\rangle +|-\rangle
}{\sqrt{2}}, ~~
 |0_O\rangle_0=
\frac{1}{N_O}\frac{|+\rangle -|-\rangle }{\sqrt{2}},
\end{eqnarray*}
where the states $|\pm\rangle$ stand for the ground states of a harmonic
oscillator centered around $x=\pm x_0$, annihilated by $\displaystyle
\hat a_\pm= \frac{i\hat p}{2p_G}+\frac{\hat x\mp x_0}{2x_G}$, satisfying
\begin{eqnarray*} \label{0:pm}
\hat a_\pm |\pm\rangle =0,~~~~\hat a_\pm |\mp\rangle= \mp
|\mp\rangle \frac{x_0}{x_G} ,
\end{eqnarray*}
and the normalization constants are $N_O^2=1-\langle
+|-\rangle$ and $N_E^2=1+\langle +|-\rangle $.
The time-independent wavefunctions for states $|\pm\rangle$ are given by
the Gaussian wave-packet,
\begin{eqnarray} \label{wf:pm1}
\Psi_{\pm}(x)\equiv \langle x|\pm \rangle= \left({p_G \over
\hbar \pi x_G}\right)^{1/4}e^{-\frac{p_G}{2\hbar x_G}
    (x\pm x_0)^2}= \left({1\over
2 \pi x_G^2}\right)^{1/4}e^{-\frac{(x\pm x_0)^2}{4 x_G^2}
    }.
\end{eqnarray}
The overlap of the two states $|\pm \rangle $ used in defining the
normalization constant is
\begin{eqnarray*} \label{cross}
\langle +|-\rangle = e^{-\frac{p_G}{\hbar x_G} x_0^2} .
\end{eqnarray*}
Evidently, these even and odd ground states are orthogonal to each
other:
\begin{eqnarray} \label{exp}
_0\langle 0_E|0_O\rangle_0 =0 .\nn
\end{eqnarray}
We now show that the states $|0_{E/O}\rangle_0$ and an operator
($\displaystyle \hat A_0^\dagger \sim \frac{x_G}{\bar x} \hat
a_+^\dagger \hat a_-^\dagger$), where $\bar x=\sqrt{\langle \hat x^2
\rangle} $, correctly describe the two limiting Hilbert spaces, the
harmonic oscillator limit ($\omega^2
>0,~\lambda \rightarrow 0$ for $x_0\rightarrow 0$) and
the infinitely separated double-well limit ($\omega^2<0,~\lambda
\rightarrow 0$ for $x_0\rightarrow \infty$) of the system described by
the Hamiltonian~(\ref{H:ah}).

Consider the harmonic oscillator limit first.
The ground state wavefunction of a harmonic oscillator is a Gaussian.
Since the sum of two Gaussian wave packets of the same frequency is a
Gaussian, the ground state wavefunction is correctly reproduced in the
$x_0\rightarrow 0$ limit.
Let us examine the $1^{st}$ excited state in the $x_0 \rightarrow 0$
limit.
The odd parity ground state becomes an exact first excited state
wavefunction of a harmonic oscillator in the $x_0=0$ limit:
\begin{eqnarray*}
\lim_{x_0 \rightarrow 0}\langle x|0_O\rangle_0 =
    \left(\frac{4 p_G}{\hbar_G \pi x_G}\right)^{1/4}
     \sqrt{\frac{p_G}{\hbar x_G}} x \exp\left(
     -\frac{p_G x^2}{2\hbar x_G}
    \right).
\end{eqnarray*}
Therefore $|0_{E/O}\rangle $ gives the exact ground state and the first
excited state in the harmonic oscillator limit.
Repeated actions of the operator $\hat A_0^\dagger \sim \hat a_+^\dagger
\hat a_-^\dagger= \hat a^{\dagger 2}$, with $x_G=\bar x$, on the two
states $|0_{E/O}\rangle_0 $ reproduce the complete Hilbert space of the
harmonic oscillator.
Both of the states $|0_{E/O}\rangle_0 $ are annihilated by $\hat A_0\sim
\hat a_+ \hat a_-=\hat a^2$.
In this sense, the set of operators ($\hat A_0$, $\hat A_0^\dagger$) and
the even and odd ground states $|0_{E/O}\rangle$ correctly reproduce the
harmonic oscillator system in the $x_0 \rightarrow 0$ limit and $\hat
A_0$ and $\hat A_0^\dagger$ play the role of an annihilation operator
and a creation operator which raises and lowers the states by two steps.

We next consider the infinitely separated double-well limit
by setting $x_0 \rightarrow \infty$.
Now, the two states $|\pm\rangle$ are completely separated from each
other and do not overlap.
Therefore, the energy of the two states $|0_{E/O}\rangle_0 $ are
degenerated and the operator $\hat A_0^\dagger$ acts separately on each
states $|\pm \rangle$ and raises each states $|\pm\rangle$ to the
$1^{st}$ excited states.
For example, for $|+\rangle$ state,
\begin{eqnarray*}
\frac{x_G}{\bar x} \hat a_+^\dagger \hat a_-^\dagger
|+\rangle = \frac{x_G}{\bar x}\hat a_+^\dagger \left(\hat
a_+^\dagger +\frac{x_0}{x_G}\right)|+\rangle =
\frac{x_0}{\bar x} \hat a_+^\dagger |+\rangle
+O(\frac{x_G}{\bar x})\Rightarrow |1_+\rangle ,
\end{eqnarray*}
where in the infinitely separated double-well limit,
$x_0/\bar x\rightarrow 1$ and $x_G/\bar x\rightarrow 0$.
Similar calculations will be applicable to the state $|-\rangle$.
In addition, $\hat A_0$ annihilates both of the states $|\pm \rangle$,
simultaneously.
In this sense, the operators $\hat A_0$ and $\hat A_0^\dagger$ play the
role of an annihilation operator and a creation operator and correctly
produce the Hilbert space of the infinitely separated double-well
oscillator in the $x_0\rightarrow \infty$ limit.

This inspection of the two limits $x_0 \rightarrow 0,~
\infty$ implies that, by using the even power operators
$\hat A_0$ and $\hat A_0^\dagger$, one may describe both
the anharmonic and the double-well oscillators
simultaneously.

We construct a $0^{th}$ order invariant annihilation operator $\cA$ of
the exact annihilation operator $\hat A$ of the system~(\ref{H:ah}) in
Sec. II by solving the LvN equation to the lowest order for variational
approximation.
We construct the creation ($\cA^\dagger$) and annihilation ($\cA$)
operators as even-power series in $\hat p$ and $\hat x$ in such a way
that they raise or lower the energy eigenstates by two steps.
For example, the creation operator raises the ground state to the
$2^{nd}$ excited state and the $1^{st}$ excited state to the $3^{rd}$
excited state, etc.
The annihilation operator $\cA$ must annihilate both of the ground state
$|0_E\rangle\equiv |0\rangle$ and the first excited state
$|0_O\rangle\equiv |1\rangle$, and their wavefunctions are an even
function and an odd function of $x$, respectively.
In the harmonic oscillator limit, $x_0\rightarrow 0$, the annihilation
operator must become the square of the annihilation operator of the
harmonic oscillator.
In the infinitely separated double-well limit, $x_0\rightarrow \infty$,
the annihilation operator $\cA$ must reproduce the annihilation operator
of a harmonic oscillator which is centered at each bottom of the
double-well potential.
An algebra satisfied by the operators are obtained and then
used to construct the Hilbert spaces for the oscillators.
This operator $\cA$ is used in Sec. III to construct the variational
approximations of the anharmonic and the double-well oscillators.
It is shown to provide a good approximation for the
energies of the oscillator eigenstates with errors smaller
than $10^{-2}$\% for most range of parameters of both the
double-well oscillator and the anharmonic oscillator.
We then calculate the first order correction to the invariant operators
$(\cA,\cA^\dagger)$, and a systematic method to construct higher order
invariant operators in perturbative series is developed in Sec. IV.
We conclude with some discussions on our method in the last
section.

\section{Zeroth order solution of the Liouville-von Neumann Equation}

The variational perturbation method starts from variationally
identifying the $0^{th}$ order approximate ground state and an
annihilation operator which annihilates the ground state.
For example, the Gaussian approximation uses the Gaussian
wave-packet, $Ne^{-\Omega_G x^2/2}$, as the zeroth order
ground state, with $\Omega$ a variational parameter.
In this section, we try to find a $0^{th}$ order ground
state and an annihilation operator, which describe both the
double-well oscillator and the anharmonic oscillator,
consistently.

\subsection{Localization condition}
There have been attempts to find the solution $\hat a$ of
the LvN equation~(\ref{LvN}) for anharmonic oscillator as a
Taylor-like series of $\hat p$ and $\hat x$~\cite{kim}.
This process reveals that the anharmonic
oscillator~(\ref{H:ah}) with $\omega^2>0$ has the structure
of a $q-$deformed oscillator to the first order in the
variational perturbation series and that of a generalized
deformed oscillator to the next order~\cite{kimq}.
This observation is based on the fact that the low lying
eigenstates of anharmonic oscillator are localized around
$x=0$ and $p=0$, in the sense that,
\begin{eqnarray} \label{con:G0}
\langle \hat x\rangle \sim 0,~~~~
 \langle \hat x^2\rangle \sim \frac{\hbar}{2\Omega},
 ~~~~\langle \hat p\rangle \sim 0,~~~~
 \displaystyle \langle \hat p^2\rangle \sim
 \frac{\hbar \Omega}{2}.
\end{eqnarray}
In the Gaussian approximation, this condition is implicitly
assumed.
But for the double-well systems the
conditions~(\ref{con:G0}) are not satisfied.

The probability distributions of low-lying eigenstates of a
double-well oscillator would be bi-centered around $x=\pm
x_0~(x_0\geq 0)$ with $\langle p\rangle =0$.
We thus develop a generalized series expansion in terms of
bi-centered polynomials.
For example, a non-singular function $f(x)$ can be written
as
\begin{eqnarray} \label{f:approx}
f(x) = \sum _{n=0}
\left[\frac{f^+_{n}}{(2n)!}(x^2-x_0^2)^n+
\frac{f^-_{n}}{(2n+1)!} x(x^2-x_0^2)^n\right] .
\end{eqnarray}
Identifying $f^\pm_n$ for $n=0,1,2\cdots$ provides an
approximation of $f(x)$ around $x \sim \pm x_0$ in a series
form.
If one wants to describe physics around $x=\pm x_0$, this
approximation can be used effectively.

We assume the energy scale of each packet at $x=\pm x_0$ to
be $\hbar \Omega$.
Using dimensional analysis, we have the `localization
condition',
\begin{eqnarray} \label{exp:0}
\langle \hat x\rangle \sim 0,~~\langle (x^2-
x_0^2)^2\rangle
    \sim
(2\bar x)^2\frac{ \hbar}{2\Omega},~~ \langle \hat p\rangle
    \sim 0,~~\langle p^2\rangle \sim \frac{\Omega \hbar}{2},
\end{eqnarray}
where $\bar{x}=\sqrt{\langle \hat x^2 \rangle }$.
In this condition, three different length scales appear:
the position of each bi-centered packets ($x_0$), the
root-mean-square expectation value of the position operator
($\bar x$), and the length scale corresponding to the
energy scale ($x_\Omega=\sqrt{\hbar/(2\Omega)}$).
The ratios of the three scales provide interesting
dimensionless non-negative quantities which are not larger
than 1,
\begin{eqnarray} \label{e:O x}
 \epsilon\equiv \frac{x_\Omega}{\bar{x}}
    =\sqrt{\frac{\hbar}{2\Omega
    \bar{x}^2}} \leq 1
  ,~~ \varepsilon \equiv \frac{x_0^2}{\bar x^2} \leq 1 .
\end{eqnarray}
The inequalities can be easily proven for the wave-packet
given by the sum of two equal packets:
\begin{eqnarray*} \label{wave}
\psi_\pm(x) = \frac{1}{\sqrt{2N}}\left[\phi(x-x_0)\pm
\phi(x+x_0)\right] ; ~~~~N= 1\pm
\int_{-\infty}^{\infty}dx'\phi(x'-x_0)\phi(x'+x_0) .
\end{eqnarray*}
With respect to this packet, we have the relation
\begin{eqnarray*} \label{rel:size}
\bar x^2= x_\Omega^2+\frac{x_0^2}{N} ,
\end{eqnarray*}
where we have assumed the wave-function being real for
simplicity.
For $\psi_-$ case, Eq.~(\ref{e:O x}) is evident. For
$\psi_+$ case, the value $\varepsilon=x_0^2/\bar x^2$ will
be maximized at $x_0\rightarrow \infty$, where
$\varepsilon=1$.
Therefore, the inequality in Eq.~(\ref{e:O x}) is valid.

We consider an explicit example of a bi-centered packet
given by the sum of two Gaussian packets centered at $x=\pm
x_0$, respectively.
Then we have the inequality $\bar x \geq x_0$ and $\bar x
\geq x_\Omega$ always, where the equalities hold for
$x_0=\infty$ and $x_0=0$, respectively.
Thus the ratios $\epsilon$ and $\varepsilon$ are always not
larger than $1$ and they take the values
$\epsilon=1(\varepsilon=0)$ for $x_0=0$ and
$\epsilon=0(\varepsilon=1)$ for $x_0 \rightarrow \infty$.
Since the multiplication of these parameters,
\begin{eqnarray} \label{delta}
\delta\equiv \epsilon \varepsilon ,
\end{eqnarray}
is always smaller than $1$, it can be used as a good
perturbation parameter for systems with bi-centered wave
packets as approximate eigenstates.
Note that these parameters are not fixed, but depend on the
nature of wavefunctions.
Thus we will use these parameters as variational
parameters.

\begin{figure}[htbp]
\begin{center}
\includegraphics[width=.45\linewidth,origin=tl]{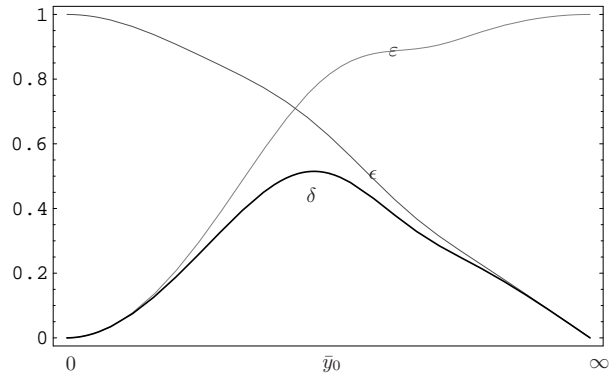}
\end{center}
\caption{Schematic plot of the values of $\epsilon$, $\varepsilon$, and
$\delta$ as functions of the distance between the central positions of
the two Gaussian packets.
This is the explicit zeroth order result for even-parity eigenstates.
The horizontal distance in the figure is rescaled by the tangent
function, $\bar y_0=\tan z$. Note that the explicit numerical value of
$\delta$ is smaller than $1/2$. } \label{fig1}
\end{figure}

For calculational convenience we introduce dimensionless
operators $\hat \pi$ and $\hat y$ for momentum and position
operators defined by
\begin{eqnarray} \label{var:dimless}
\hat \pi= \frac{\hat p}{\sqrt{\hbar \Omega}}, ~~
 \hat y=\frac{\hat x}{\sqrt{\hbar/\Omega}} .
\end{eqnarray}
Note also that $y_0=x_0\sqrt{\Omega/\hbar}$ relates the two
parameters $\epsilon$ and $\varepsilon$ by
\begin{eqnarray} \label{ep:vep1}
 2\epsilon^2 y_0^2= \varepsilon .
\end{eqnarray}
 The commutation relation between $\hat y$ and $\hat \pi$
is
\begin{eqnarray} \label{ypi:comm}
[\hat y, \hat \pi] &=& i ,
\end{eqnarray}
and the localization condition~(\ref{exp:0}) becomes
\begin{eqnarray} \label{exp:ypi}
\langle \hat y\rangle \sim 0,~~\langle (\hat
y^2-y_0^2)^2\rangle \sim \frac{1}{\epsilon^2} , ~~ \langle
\hat \pi \rangle \sim 0,~~\langle \hat \pi^2 \rangle \sim
\frac{1}{2}.
\end{eqnarray}
The expectation value of $\hat y^2$ is explicitly given by
\begin{eqnarray} \label{y2:exp}
\langle y^2\rangle = \frac{\langle \hat
x^2\rangle}{2\hbar/(2\Omega)}= \frac{\bar
x^2}{2x_\Omega^2}=\frac{1}{2\epsilon^2} .
\end{eqnarray}

For later use, we introduce some operators of
$O(\epsilon^0)$ and $O(\varepsilon^0)$,
\begin{eqnarray} \label{ex}
T_1^0=\sqrt{2} \hat \pi,~~ T_0^1= \sqrt{2}\epsilon \hat
y,~~ T_2^0=2 \hat \pi^2, ~~ T_0^2=\epsilon(\hat
y^2-y_0^2),~~
    T_1^1=\epsilon\{\hat \pi, \hat y\}_+
    ,~~ T_0^4= \epsilon^2(\hat y^2-y_0^2)^2,
\end{eqnarray}
which will be useful in describing the bi-centered systems.

\subsection{Hamiltonian for variational approximation }
We now consider a quartic oscillator, which will be used as
a basis for the variational approximation, described by the
Hamiltonian:
\begin{eqnarray} \label{H:sim}
H_0=\frac{\hbar \Omega}{2}H_m+ V_0;~~~~H_m
=\frac{1}{2}T_2^0 +
    \frac{v_2 \epsilon}{\epsilon_0^2} T_0^2 +
    \frac{\bar \lambda \varepsilon}{2} T_0^4,
\end{eqnarray}
where $V_0$ is the potential value at $\hat y=y_0$ and the
subscript in $H_0$ denotes that this Hamiltonian will be
used as a $0^{th}$ order form for the variational
perturbative expansion of double-well and anharmonic
oscillators.
The constant $\epsilon_0$ is the value of $\epsilon$ in the harmonic
oscillator limit, $y_0=0$.
Explicitly, $\epsilon_0=1$ for $|0_E\rangle_0$ and
$\epsilon_0=1/\sqrt{3}$ for $|0_O\rangle_0$.
Note that this is one of the most general form of quartic Hamiltonian if
$\Omega$ and $y_0$ are chosen arbitrarily.
Note also that only one of the three quantities $\epsilon$,
$\varepsilon$, and $y_0$ is an independent variable because of
Eqs.~(\ref{e:O x}), (\ref{ep:vep1}), and (\ref{y2:exp}).
Since we are to use the ground states of~(\ref{H:sim}) as trial
wavefunctions for the variational approximation for the
Hamiltonian~(\ref{H:ah}), we have two variational parameters, $\Omega$
and $y_0$.

The coefficients ($v_2$, $\bar\lambda$) in the
Hamiltonian~(\ref{H:sim}) can be written in terms of the
above mentioned parameters by considering the two limiting
cases, the harmonic oscillator ($y_0=0$) and the
infinitely-separated ($y_0\rightarrow \infty$) double-well
limits.

In the harmonic oscillator limit, the ground state is given
by a Gaussian packet.
The exact frequency is given by $\Omega=\omega$ and there is no quartic
potential in the Hamiltonian.
Therefore, the Hamiltonian~(\ref{H:sim}) becomes that of the harmonic
oscillator with correct frequency $\omega$ if
\begin{eqnarray*}
\mbox{harmonic oscillator limit} ~(\varepsilon \rightarrow 0,~~\epsilon
\rightarrow \epsilon_0):\left\{\begin{tabular}{l}
$v_2(y_0\rightarrow 0)= 1,$\\
$\bar \lambda (y_0\rightarrow 0)=\mbox{finite number}$.\\
\end{tabular}\right.
\end{eqnarray*}
On the other hand, for the ground state of the infinitely
separated double-well limit ($y_0\rightarrow \infty$), we
have $\epsilon=0$ and $\varepsilon=1$ and the $T_0^2$ term
disappears since $\epsilon$ vanishes.
Each packet at $x=\pm x_0$ evolves as if it is a free harmonic
oscillator.
The frequency of each packets is determined by the second derivative of
the potential with respect to $x$ at the bottom of the potential.
This determines $\Omega^2=2\lambda x_0^2=2\lambda \bar x^2$.
Therefore, the coefficient $\bar \lambda$ of $T_0^4$ term must be $1$ in
this limit:
\begin{eqnarray*}
 \mbox{infinitely separated double-well
limit}~(\varepsilon \rightarrow 1,~~ \epsilon \rightarrow
0):\left\{\begin{tabular}{l}
$v_2(y_0\rightarrow \infty)=\mbox{finite number},$\\
$\bar\lambda (y_0\rightarrow \infty)=1$.\\
\end{tabular}\right.
\end{eqnarray*}
With these observations, it is clear that the constants $v_2$ and $\bar
\lambda$ are of $O(\delta^0)$.
The values $v_2$ and $\bar \lambda$ become important at the harmonic
oscillator limit and the infinitely separated double-well limit,
respectively.
For both limits, therefore, we may set
\begin{eqnarray} \label{coef}
v_2=1= \bar \lambda,
\end{eqnarray}
for the zeroth order approximate Hamiltonian $H_0$.
With the values of $v_2$ and $\bar \lambda$ of Eq.~(\ref{coef}), the
Hamiltonian~(\ref{H:sim}) still describes the most general anharmonic
and double-well oscillators since $\Omega$ and $y_0$ are free
parameters.

\subsection{Zeroth order results}

As we have discussed in the previous section, we try to
find an annihilation operator $\hat A$ and a creation
operator $\hat A^\dagger$ to $O(\delta^0)$ in this
subsection.
Since we want to find the operators as even functions of the bi-centered
operators we try the simplest ansatz for the zeroth order operator
$\cA$:
\begin{eqnarray} \label{A:anhar}
\cA&=& u(t) T_0^2 +v(t) T_1^1+ w(t) T_2^0. \nn
\end{eqnarray}
To the zeroth order, one can easily find that the LvN
equation, $\dsp i \hbar \frac{d \cA(\hat p,\hat x,t)}{dt} =
i \hbar \frac{\partial
    \cA}{\partial t}+[\cA,H_0]=0$,
 becomes a matrix differential equation of the form
\begin{eqnarray*} \label{diff:LvN}
\frac{d {\bf A}(t)}{dt}= \frac{i \Omega}{2} M {\bf A}(t) ,
~~~~
 {\bf A}(t)=\left(\begin{tabular}{c}
    $u$ \\ $v$ \\ $w$
        \end{tabular}\right),
\end{eqnarray*}
where $M$ is a $(3\times 3)$ matrix determined by the
commutator of $\cA$ and $H_m$ of Eq.~(\ref{H:sim}).
For the time-independent case, the differential equation can be solved
by a simple matrix eigenvalue equation by setting $\displaystyle {\bf
A}(t)= e^{i \zeta \Omega t} {\bf A}(t=0)$.
Explicit calculation of the commutator $[\cA, H_m]$ gives the matrix
$M$:
\begin{eqnarray*} \label{ab:eq0}
M=-i \left(\begin{tabular}{ccc}
            0 & 0 & $2 \epsilon^2/\epsilon_0^2+\varepsilon^2$ \\
            0 & 0 & $-\epsilon$ \\
            $-1$ & $2\epsilon/\epsilon_0^2 $& 0 \\
            \end{tabular}\right) .
\end{eqnarray*}
The eigenvalues of $M$ are given by the solution $\zeta$ of
\begin{eqnarray*} \label{M:eigen}
 -\zeta (\zeta^2 - \alpha^2)=0,
\end{eqnarray*}
where
\begin{eqnarray} \label{alpha}
\alpha=\sqrt{\frac{4\epsilon^2}{\epsilon_0^2}+\varepsilon^2}
.
\end{eqnarray}
We have three different $\zeta$ eigenvalues, $(0, \pm
\alpha)$.
The $\zeta=0$ solution is, in fact, the Hamiltonian itself.
Since what we want to find in this section is the creation and the
annihilation operators, we do not try to analyze it here.

The other two solutions with eigenvalues $\zeta=\pm \alpha$
denote, in fact, a set of operators $\cA$ and
$\cA^\dagger$.
If we demand $\cA$ being a negative frequency mode, we have
\begin{eqnarray} \label{ab:sol0}
\cA&\equiv&  b_0 e^{i\alpha \Omega t}\ca\,;\quad\quad \ca=
-i\nu^2 T_0^2+T_1^1+\frac{i\epsilon}{\alpha} T_2^0 ,\\
\cA^\dagger &\equiv&  b_0 e^{-i\alpha \Omega
    t}\ca^\dagger\,;\quad\quad \ca^\dagger= i\nu^2
T_0^2+T_1^1-\frac{i\epsilon}{\alpha} T_2^0 ,\nn
\end{eqnarray}
where $b_0$ is a normalization constant to be determined
later and $\nu$ is
\begin{eqnarray} \label{nu}
\nu^2=\frac{2\epsilon^2/\epsilon_0^2+\varepsilon^2}{\alpha}
.
\end{eqnarray}
These are the zeroth order annihilation and creation operators we were
looking for. Thus the operator $\cA$ satisfies
\begin{eqnarray*} \label{diff}
\frac{d \cA(t,\hat p(t),\hat x(t))}{dt} \simeq 0 ,
\end{eqnarray*}
where $\simeq$ denotes "the same up to $O(\delta^0)$" and $\cA$ and
$\cA^\dagger$ are the zeroth order invariant operators. The explicit
algebra satisfied by the operators $\cA$ and $\cA^\dagger$ will be shown
in the next subsection.

\subsection{Algebraic properties of the zeroth order invariant
     operators}
In the previous subsection, we obtained the annihilation and the
creation operators which satisfy the zeroth order LvN equation. We
expect that the zeroth order solution of the LvN equation has the
structure of a generalized deformed oscillator as we have pointed out in
the introduction. In this subsection, we show that there exist well
defined algebraic properties of the operators $\cA$ and $\cA^\dagger$.

Using Eqs.~(\ref{ex}) and (\ref{ypi:comm}), we find the
commutator of the operators $\cA$ and $\cA^\dagger$:
\begin{eqnarray} \label{comm0}
\frac{[\cA, \cA^\dagger]}{4 b_0^2}&=& \nu^2
    \varepsilon
     +\frac{4 \epsilon^2 H_m}{\alpha}+\frac{2\delta}{\alpha}
     \left(\varepsilon T_0^2
        -\epsilon T_0^4 \right)
         \\
 &= &  \alpha \varepsilon
     +\frac{4 \epsilon^2 H_m}{\alpha}+ O(\delta) .\nn
\end{eqnarray}
The anticommutator of the operators $\cA$ and $\cA^\dagger$
is
\begin{eqnarray}
\frac{\{\cA,\cA^\dagger\}_+}{4 b_0^2}
 &=&\frac{3\epsilon^2}{2}+
   \varepsilon \left(\frac{1}{2}T_2^0
    +\frac{\varepsilon}{2}T_0^4\right)
    +\frac{2\epsilon^2}{\alpha^2}
    \left(\frac{1}{2}T_2^0
    +\frac{\epsilon}{\epsilon_0^2}T_0^2\right)^2
     \label{anticom0}
 \\
 &= &\varepsilon H_m
        +\frac{\epsilon^2}{2}\left(3+\frac{4H_m^2}{\alpha^2}
        \right)+O(\delta)
         .\nonumber
\end{eqnarray}

The algebraic structure of the deformed oscillator will be
apparent by introducing a structure function $\Phi_0(\hat
N_0)$ of the form
\begin{eqnarray} \label{phi0}
\Phi_0(\hat N_0) &=& 4 b_0^2\left({\cal E}_+
 \hat N_0+4\epsilon^2\hat N_0^2\right).
\end{eqnarray}
By comparing Eqs.~(\ref{comm0}), (\ref{anticom0}), and
(\ref{phi0}) with Eq.~(\ref{gen:osc}), we have
\begin{eqnarray} \label{cE}
{\cal E}_+ &=&\alpha \varepsilon + 4\epsilon^2
\left(\frac{\cE_0}{\alpha}-1\right)  .\nn
\end{eqnarray}
The constant $\cE_0$ is determined by the condition~$\Phi_0(0)=0$. With
(\ref{comm0}) and (\ref{anticom0}), the condition reads
\begin{eqnarray} \label{E0}
\alpha\varepsilon
\left(\frac{\cE_0}{\alpha}-1\right)+2\epsilon^2
    \left[\left(
    \frac{\cE_0}{\alpha}-1\right)^2-\frac{1}{4}\right]
    \simeq 0 .
\end{eqnarray}
This equation determines the energy eigenvalues in both the
harmonic oscillator and the infinitely separated
double-well limits.
For the infinitely separated double-well limit ($y_0\rightarrow
\infty,~\epsilon \rightarrow 0$), we have $\cE_0=\alpha$ and for the
harmonic oscillator limit ($y_0\rightarrow 0,~\varepsilon \rightarrow
0$), we have two different solution $\cE_0= 3\alpha/2,~ \alpha/2$.
These two eigenvalues correspond to the energy eigenvalues of the
odd-parity and the even-parity ground states.
The rescaled ground state energy $E_0$ in Table 1, which is determined
by Eq.~(\ref{E:1}) below, satisfies this condition~(\ref{E0}).

The structure function~(\ref{phi0}) and the number operator
$\hat N_0$,
\begin{eqnarray} \label{N}
\hat N_0 = \frac{H_m-\cE_0}{2\alpha},
\end{eqnarray}
satisfy the conditions for the generalized deformed
oscillator, Eq.~(\ref{gen:osc}), since
\begin{eqnarray} \label{ada}
\cA^\dagger \cA \simeq \Phi_0(\hat N_0),~~
 \cA \cA^\dagger \simeq \Phi_0(
    \hat N_0+1) .
\end{eqnarray}
The Hamiltonian can be written in terms of the number
operator $\hat N_0$ as
\begin{eqnarray} \label{H:N}
H \simeq \frac{\hbar \Omega}{2}(2\alpha \hat
N_0+\cE_0)+V(x_0) .
\end{eqnarray}
Therefore, the ground state energy is given by $\dsp
\frac{\hbar \Omega}{2}\cE_0+V(x_0)$ and the energy
difference between two nearby states in each parity sector
is $\alpha \hbar \Omega$.

The yet undetermined quantity $b_0$  in Eq.~(\ref{ab:sol0})
is fixed by imposing $[\cA, \cA^\dagger]|0_{E/O}\rangle_0=
|0_{E/O}\rangle_0 $. This gives
\begin{eqnarray} \label{b:0}
\Phi_0(1) \equiv 1~~~ \Longrightarrow ~~~
b_0^2=\frac{1}{4({\cal E}_++4\epsilon^2) } .
\end{eqnarray}
Note also that  due to the LvN equation~(\ref{LvN}), (\ref{H:sim}), and
the definition of $\hat N_0$~(\ref{N}), we have the desired commutation
relation:
\begin{eqnarray} \label{Na:com}
~[\cA,\hat N_0]\simeq \cA ,\quad
 ~~[\cA^\dagger,\hat N_0]\simeq
    -\cA^\dagger .
\end{eqnarray}
With these algebra and Eq.~(\ref{phi0}), the ground state
$|0_{E/O}\rangle_0$ can be defined as an eigenstate of
$\hat N_0$:
\begin{eqnarray} \label{0:A0}
 \cA |0_{E/O}\rangle_0 \simeq 0.
\end{eqnarray}
From Eq.~(\ref{ada}) and the condition $\Phi_0(0)=0$, the
number operator annihilates the ground states:
\begin{eqnarray} \label{..}
\hat N_0|0_{E/O}\rangle_0 \equiv |0_{E/O}\rangle_0
\Phi_0^{-1}(0)=0 .
\end{eqnarray}
The $n^{th}$ excited states are generated by the formula:
\begin{eqnarray} \label{n:da}
|n_{E/O}\rangle_0 = \frac{(\hat
    {\cal A}^\dagger)^n}{\sqrt{[n]_0!}} |0_{E/O}
    \rangle_0 ,
\end{eqnarray}
where
\begin{eqnarray} \label{[n]}
[n]_0!\equiv \prod^n_{k=1}[k]_0= \prod_{k=1}^n \Phi_0(k) .
\end{eqnarray}
The states $|n_{E}\rangle_0$ and $|n_{O}\rangle_0$ are the $(2n)^{th}$
and the $(2n+1)^{th}$ excited states with even and odd parities of the
anharmonic and the double-well oscillator. Due to Eqs.~(\ref{ada}) and
(\ref{Na:com}), the states have the eigenvalues $n_{E/O}$ for $\hat
N_0$:
\begin{eqnarray} \label{eigenstates}
\hat N_0|n_{E/O}\rangle_0 \simeq n_{E/O}|n_{E/O}\rangle_0 .
\end{eqnarray}
Moreover, these eigenstates are orthonormal to each other
to the zeroth order:
\begin{eqnarray} \label{ortho}
_0\langle n_{E/O} |m_{E/O}\rangle_0 \simeq \delta_{nm} .
\end{eqnarray}
Eqs.~(\ref{phi0}), (\ref{N}),  (\ref{ada}), (\ref{Na:com}), and
(\ref{0:A0}) constitute the conditions for a generalized deformed
oscillator which possesses a Fock space of eigenvectors
$|0_{E/O}\rangle_0,~|1_{E/O}\rangle_0,~\cdots, |n_{E/O}\rangle_0,
\cdots$ of the number operator $\hat N_0$.
Further specification of $\hat A$ to higher order is possible
perturbatively, which is the purpose of Sec. IV.

\section{Variational approximation}
In the previous section we have obtained a zeroth order invariant
creation and annihilation operators which satisfies the LvN equation for
the zeroth order Hamiltonian~(\ref{H:sim}) which depends on two
arbitrary parameters, $\Omega$ and $y_0$.
By using the ground state of the zeroth order Hamiltonian~(\ref{H:sim})
as a trial wavefunction we develop, in this section, a variational
approximation of the anharmonic and the double-well oscillators.
We then compare our result with those of other approximation methods.
%

In coordinate space $y$, the eigenstate $\Psi(y)$, which satisfies $\cA
\Psi(y)=0$, is given by the linear combination of $e^{-\alpha y^2}
H_m(\beta y)$ and the confluent hypergeometric function $_1F_1$, which
give difficulty in analytic manipulations.
Therefore, we propose to use an operator $\hat a_0$ which is equivalent
to $\cA$ up to the zeroth order but have simpler ground state:
\begin{eqnarray*} \label{equi:double}
\hat a_0 \equiv -2i \epsilon \hat a_+ \hat a_-
 \simeq -2i\epsilon \hat a_+ \hat
    a_- +   \frac{
    i\varepsilon \delta
    }{2\nu^2 \alpha^2 }T_2^0 =\frac{\cA}{b_0e^{i\alpha\Omega t}}= \ca,
\end{eqnarray*}
where
\begin{eqnarray} \label{a:pm} \hat a_\pm
=\frac{i}{\sqrt{2}\nu } \hat \pi+ \frac{\nu}{\sqrt{2}}(\hat
y\mp y_0)
\end{eqnarray}
are the annihilation operators of harmonic oscillators
centered around $y=\pm y_0$. The new operator is explicitly
written as
\begin{eqnarray} \label{a0:def}
\hat a_0&=&
    -i\nu^2 T_0^2
    +T_1^1+\frac{i\epsilon}{2\nu^2}T_2^0
  .
\end{eqnarray}
The operators $\hat \pi$ and $\hat y$ are related to $\hat
a_\pm$ by
\begin{eqnarray} \label{ops}
\hat \pi&=&-\frac{i\nu}{\sqrt{2}}(\hat a_\pm -\hat a_\pm ^\dagger), \\
-\sqrt{2}\nu y_0 &=&\hat a_+-\hat a_-, \nonumber \\
\hat y&=&\pm y_0 + \frac{\hat a_\pm +\hat a_\pm
    ^\dagger}{\sqrt{2}\nu} . \nonumber
\end{eqnarray}
Thus $\cA$ is equivalent to the zeroth order to the product of the two
annihilation operators of harmonic oscillators centered around $y=\pm
y_0$.
Note that the annihilation operators $\hat a_\pm$ and the creation
operators $\hat a_\pm^\dagger$ satisfy the standard commutation relation
\begin{eqnarray} \label{comm:apm}
[\hat a_\pm, \hat a_\pm^\dagger]=1, \quad\quad [\hat
a_+,\hat a_-]=0.
\end{eqnarray}
Consider the states $|\pm\rangle$ annihilated by $\hat
a_\pm$,
\begin{eqnarray} \label{0:pm}
\hat a_\pm |\pm\rangle =0 .
\end{eqnarray}
The wave-functions for states $|\pm\rangle$ are
\begin{eqnarray*} \label{wf:pm}
\Psi_{\pm}(y)\equiv \langle y|\pm \rangle= {\nu^{1/2} \over
\pi^{1/4}}e^{-\nu^2 (y-y_0)^2/2} ,
\end{eqnarray*}
which determine the overlap $\langle +|-\rangle$:
\begin{eqnarray*} \label{cross}
\langle +|-\rangle = e^{-\nu^2 y_0^2} .
\end{eqnarray*}
Since both of the two states $|\pm\rangle$ are annihilated by $\hat
a_0$, the approximate ground states of even or odd parity,
$|0_{E/O}\rangle$, which are annihilated by $\hat a_0$, are given by the
linear combinations,
\begin{eqnarray} \label{vac}
|0_E\rangle_0= \frac{1}{M_E}\frac{|+\rangle +|-\rangle
}{\sqrt{2}}, ~~
 |0_O\rangle_0=
\frac{1}{M_O}\frac{|+\rangle -|-\rangle }{\sqrt{2}}.
\end{eqnarray}
The normalization constants are
\begin{eqnarray} \label{Ns}
M_E^2 =1+\langle +|-\rangle= 1 +e^{-\nu^2 y_0^2} ,
\quad\quad
 M_O^2=1-\langle +|-\rangle = 1-e^{-\nu^2 y_0^2} .
\end{eqnarray}
The ground states~(\ref{vac}) satisfy
\begin{eqnarray*} \label{exp}
_0\langle n_{E/O}|0_{E/O}\rangle_0 =0,\quad \mbox{ for }
n\neq 0 ,
\end{eqnarray*}
where the excited states $|n_{E/O}\rangle_0$ are generated
by the successive action of $\hat a_0^\dagger$ on the
ground states. We thus have constructed the set of
operators $(\hat a_0, \hat a_0^\dagger)$ and the even and
odd ground states $(|0_E\rangle_0, |0_O\rangle)$, which
properly describe both the anharmonic and double-well
oscillator simultaneously to the zeroth order in $\delta$.

The expectation values of operators with respect to the
states $|0_{E/O}\rangle_0$ can be obtained algebraically by
using Eqs.~(\ref{ops}), (\ref{comm:apm}), and~(\ref{0:pm}).
For example, the expectation value of $T_2^0$ is
\begin{eqnarray} \label{T20:0}
_0\langle 0_{E/O}| T_2^0|0_{E/O}\rangle_0 &=&
-\frac{1}{2N^2}(\langle +|\pm
    \langle
    -|)(\hat a_+-\hat a_+^\dagger)^2(|+\rangle\pm |-\rangle )
    =\nu^2 T_{E/O} ,
\end{eqnarray}
where upper (lower) sign refers to the even (odd) parity case and
\begin{eqnarray} \label{T:def}
T_{E/O}(\nu y_0) &\equiv& 1-2\nu^2 y_0^2 C_{E/O}=1\mp
\frac{2\nu^2y_0^2}{e^{\nu^2 y_0^2}\pm 1}.
\end{eqnarray}

Starting from $1$, $T_E$ decreases to $0.45$ at $\nu y_0=1.1$ and then
increases again to $1$ as $\nu y_0$ increases. On the other hand, $T_O$
continuously decreases from $3$ to $1$ as $\nu y_0$ increases. The
functions $C_{E/O}$ are
\begin{eqnarray}
 C_E(\nu
y_0)&=&\langle
    +|-\rangle/M_{E}^2=(e^{\nu^2y_0^2}+1)^{-1},\nn \\
C_O(\nu y_0)&=&-\langle
    +|-\rangle/M_O^2=-(e^{\nu^2y_0^2}-1)^{-1} \nn .
\end{eqnarray}
As shown in this equation, the expectation values for the even and the
odd ground states take similar form except for the indices in $C_{E/O}$.
From now on, $C$ will denote both of $C_{E/O}$ if not specified. This
convention is applied to all other variables such as $\epsilon$, $\nu$,
$M$, $E_0$, $T$, and $R$ defined below. The expectation value of $T_0^2$
is
\begin{eqnarray} \label{T02:0}
_0\langle 0_{E/O} |T_0^2 |0_{E/O} \rangle_0 &=&
\frac{\epsilon}{2\nu^2 } T \,.
\end{eqnarray}

The formal identity $\bar x^2\equiv
\langle(x^2-x_0^2)\rangle +x_0^2$ leads to the following
relation among parameters,
\begin{eqnarray} \label{vep:ep}
\frac{2\nu^2 y_0^2}{M^2}+1=\frac{(\nu y_0)^2}{(\epsilon
y_0)^2} .
\end{eqnarray}
Since the normalization constant $M$ is a function of $\nu
y_0$, this equation determines $\epsilon y_0$ in terms of
$\nu y_0$ or vice versa.
This equation has interesting implications on the limiting cases.
In the limit $y_0 \rightarrow 0$, Eq.~(\ref{vep:ep}) determines
$\epsilon_E=1$ and $\epsilon_O=1/\sqrt{3}$.
On the other hand, in the limit $y_0\rightarrow \infty$, we have
$\epsilon \simeq 1/(\sqrt{2}y_0)$ irrespective of the parity of the
states.
The equation~(\ref{vep:ep}) can be rewritten for $\varepsilon$ as a
function of $\bar y_0=\nu y_0$:
\begin{eqnarray} \label{vep:bars}
\varepsilon(\bar y_0)= 2\epsilon^2 y_0^2=\frac{2\bar
y_0^2}{1+\frac{2\bar y_0^2}{1\pm e^{-\bar y_0^2}}} .
\end{eqnarray}
In addition, $\epsilon$ can be written as a function of $\bar y_0$, from
Eqs.~(\ref{alpha}), (\ref{nu}), and (\ref{vep:bars}), as
\begin{eqnarray} \label{ep:bars}
\frac{\epsilon(\bar y_0)}{\epsilon_0\varepsilon(\bar
y_0)}&=& \frac{\sqrt{3}}{\sqrt{2}}\left[ -2
    +\frac{3\bar y_0^4+1}{g^{1/3}(\bar y)}
    +g^{1/3}(\bar y)\right]^{-\frac{1}{2}}; \\
 g(\bar y)&=& 1+18 \bar y_0^4+
        3 \sqrt{3} \bar y_0^2\sqrt{1+11\bar y_0^4-
        \bar y_0^8} .
        \nn
\end{eqnarray}
The explicit values of $\varepsilon$, $\epsilon$, and
$\delta=\varepsilon \epsilon$ are plotted in Fig. 1.

We also calculate the expectation value of the quartic part of the
Hamiltonian:
\begin{eqnarray} \label{T04:0}
_0\langle 0_{E/O}|T_0^4 |0_{E/O} \rangle_0 &=&
\frac{\epsilon^2}{\nu^4}\left[\frac{3+8 \nu^2 y_0^2}{4}+
 \nu^2 y_0^2\left(\nu^2 y_0^2
 -3\right)C\right] \\
 &=&\frac{\epsilon^2}{4\nu^4}\left[3+8 \nu^2 y_0^2
  \pm \frac{4\nu^2 y_0^2\left(\nu^2 y_0^2
 -3\right)}{e^{\nu^2 y_0^2}\pm 1}  \right]. \nn
\end{eqnarray}

\subsection{Effective Potential}
As in the Gaussian approximation, in which the ground state of
$\displaystyle H=\frac{p^2}{2}+\frac{\Omega^2}{2}x^2$ is chosen to be
the trial wavefunction for the Hamiltonian~(\ref{H:ah}), we use the
states~(\ref{vac}), which are the $O(\delta^0)$ even and odd ground
states of the Hamiltonian~(\ref{H:sim}), as trial wavefunctions for the
variational approximation for the Hamiltonian~(\ref{H:ah}).
Therefore, the $y_0\rightarrow 0$ limit of our approximation becomes the
Gaussian approximation, and the approximation with $y_0=
1/(\sqrt{2}\epsilon )$ limit describes a double Gaussian
approximation\cite{kovner,cooper}. For this variational approximation we
write the Hamiltonian~(\ref{H:ah}) of a general anharmonic $\omega^2>0$
or a double-well $\omega^2<0$ oscillator in terms of operators defined
in Eq.~(\ref{ex}):
\begin{eqnarray} \label{H1:2}
H&=&\frac{\hat p^2}{2}+ \frac{\omega^2}{2}\hat x^2 +
    \frac{\lambda}{4}\hat x^4
    +\frac{\omega^4}{4\lambda}\theta(-\omega^2)\\
 &=&\frac{\hbar \Omega}{2}\left[\frac{1}{2}T_2^0
 +\frac{1}{\epsilon}\left(\frac{\omega^2}{\Omega^2}
    +\frac{\lambda x_0^2}{\Omega^2}\right)T_0^2
    +\frac{\lambda \hbar}{2\Omega^3 \epsilon^2}
    T_0^4\right]+V(x_0), \nn \\
 &=&\frac{\hbar |\omega|\Omega_r}{2}\left[\frac{1}{2}T_2^0
 +\frac{1}{\epsilon}\left( \frac{\sign(\omega^2)}{\Omega_r^2}
    +\frac{2\lambda_r
    y_0^2}{\Omega_r^3}\right)T_0^2
 +\frac{\lambda_r}{\Omega_r^3 \epsilon^2}T_0^4\right]
 +V(x_0),\nn
\end{eqnarray}
where the potential $V(x_0)$ is
\begin{eqnarray}
V(x_0)&=& \frac{\omega^2x_0^2}2 + \frac{\lambda x_0^4}{
        4}+\frac{\omega^4}{4\lambda}\theta(-\omega^2)
 =\frac{\hbar |\omega|\Omega_r}{2}
        \left(\sign(\omega^2)\frac{y_0^2}{\Omega_r^2}
        +\frac{\lambda_ry_0^4}{\Omega_r^3}
        +\frac{\theta(-\omega^2)}{4\lambda_r \Omega_r}\right)
     .
\end{eqnarray}
In this equation  the theta function, $\theta(-\omega^2)$, is included
to make the lowest energy of the potential vanish for $\omega^2<0$, and
$x_0= \sqrt{\hbar /\Omega}\,y_0$, and the dimensionless parameters
$\lambda_r$ and $\Omega_r$ are defined by
\begin{eqnarray} \label{dim:r}
\lambda_r=\frac{\lambda \hbar}{2|\omega|^3}, \quad~
\Omega_r= {\Omega \over |\omega|} .
\end{eqnarray}

From Eqs.~(\ref{T20:0}), (\ref{T02:0}), and (\ref{T04:0}),
the expectation values of the Hamiltonian~(\ref{H1:2}) with
respect to the ground states $|0_E\rangle$ and
$|0_O\rangle$ in Eq.~(\ref{vac}) become
\begin{eqnarray} \label{H:exp}
\frac{2 \langle H\rangle}{\hbar |\omega|} =
 \frac{\Omega_r\nu^2 T}{2}+\sign(\omega^2)
 \frac{T+2\nu^2y_0^2}{2\Omega_r\nu^2}
 +\frac{\lambda_r}{4(\Omega_r\nu^2)^2}R(\nu y_0)
 +\frac{\theta(-\omega^2)}{4\lambda_r},
\end{eqnarray}
where we have used Eq.~(\ref{vep:ep}) to eliminate
$\epsilon$ for $\nu$, and
\begin{eqnarray} \label{R:0}
R(\nu y_0)=3+12 \nu^2 y_0^2+4\nu^4y_0^4
    -4\nu^2 y_0^2 (3+ \nu^2y_0^2) C
    =3
    + \frac{4\nu^2 y_0^2 (3+\nu^2y_0^2)}{1 \pm
        e^{-\nu^2y_0^2}}.
\end{eqnarray}
The function $R_{E}~(R_O)$ monotonically increases from $3 ~(15)$ as
$\nu y_0$ increases.
Note that the rescaling,
\begin{eqnarray} \label{rescaling}
\Omega_r \nu^2 \equiv \bar \Omega, \quad\quad \nu y_0 \equiv \bar y_0
\,,
\end{eqnarray}
completely remove the $\nu$ dependence on the expectation
values since $T$ defined by~(\ref{T:def}) is the function
only of $\nu y_0$.
The value of $\nu$ only affects the relation between $\epsilon$ and
$y_0$ through Eq.~(\ref{vep:ep}).
With these new parameters, the expectation value~(\ref{H:exp}) becomes
\begin{eqnarray} \label{H:exp2}
\frac{2 \langle H\rangle}{\hbar |\omega|} =
 \frac{\bar \Omega T(\bar y_0)}{2}+\sign(\omega^2)
 \frac{T(\bar y_0)+2\bar y_0^2}{2\bar \Omega}
 +\frac{\lambda_r}{4\bar \Omega^2}R(\bar y_0)
 +\frac{\theta(-\omega^2)}{4\lambda_r}.
\end{eqnarray}

The variation of~(\ref{H:exp2}) with respect to $\bar \Omega$ gives the
gap equation,
\begin{eqnarray} \label{dOm=0}
&&\bar \Omega^3  -\sign(\omega^2) \left(1+\frac{2\bar
y_0^2}{T(\bar y_0)}\right)
 \bar\Omega -\frac{\lambda_r R(\bar
y_0)}{T(\bar y_0)}=0 ,
\end{eqnarray}
which determines non-trivial unique positive solution for
$\bar \Omega$
\begin{eqnarray} \label{Om:sol}
\bar \Omega (\bar y_0) &=&
    \left(\frac{2}{3}\right)^{1/3}\left[ \frac{D^{1/3}}{
    3\cdot2^{1/3}}
        +\sign(\omega^2) \frac{ 1+\frac{2\bar y_0^2}{T(\bar
y_0)}}{D^{1/3}}\right];\\
D &=& \frac{9\lambda_r R(\bar y_0)}{T(\bar
    y_0)}+2\sqrt{3}\sqrt{-\sign(\omega^2)\left(1+\frac{2\bar y_0^2}{
    T(\bar y_0)}\right)^3+ \frac{27}{4}
    \frac{ \lambda_r^2R^2(\bar y_0)}{T^2(\bar y_0)}
}, \nn
\end{eqnarray}
for all values of $\omega^2$ irrespective of its sign if $\lambda \geq
0$.
We set this solution be $\bar \Omega(\bar y_0)$.
Note that the gap equation~(\ref{dOm=0}) takes similar form as that of
the variational Gaussian approximation except for the dependence of the
coefficients on $\bar y_0$.

\begin{figure}[htbp]
\begin{center}
\begin{tabular}{cc}
\includegraphics[width=.4\linewidth,origin=tl]{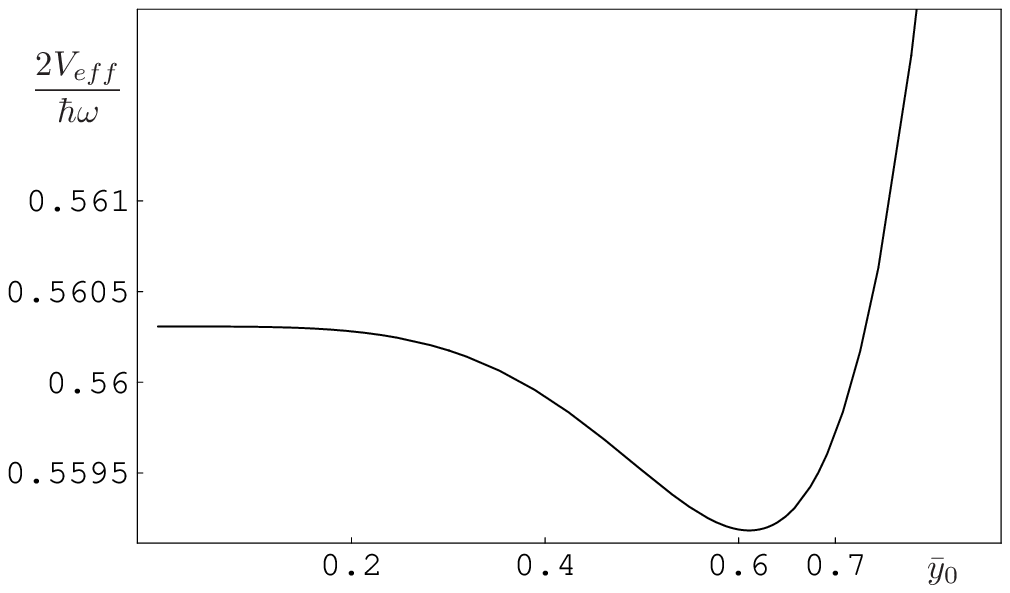} &
\includegraphics[width=.4\linewidth,origin=tl]{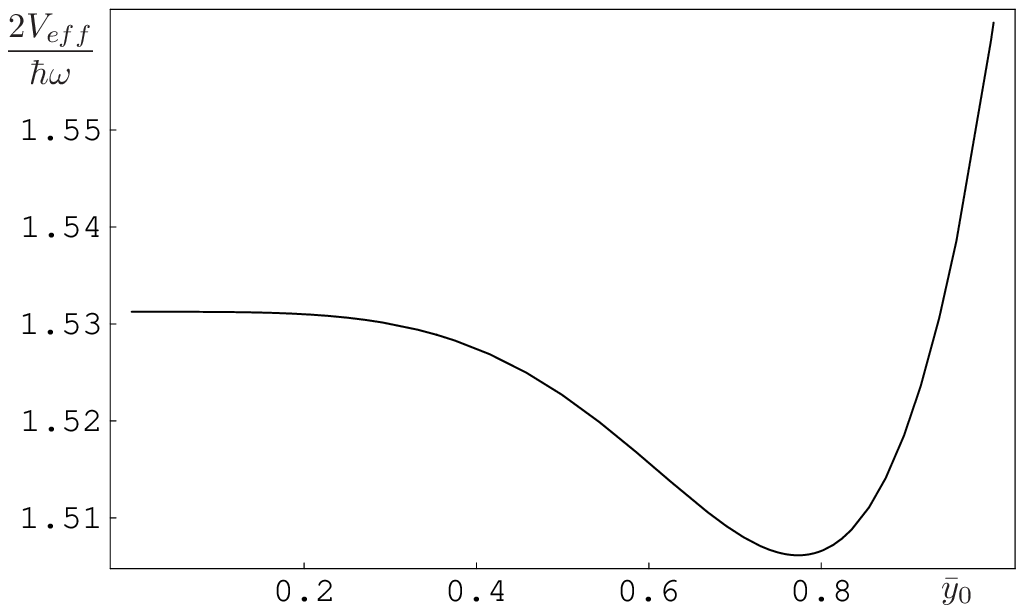}\\
\end{tabular}
\end{center}
\caption{Effective potential $\dsp \frac{2 V_{eff}(\bar
y_0)}{\hbar |\omega|}$ as a function of $\bar y_0$ for the
coupling constants
$\lambda_r= 2$ (left) and $\lambda_r=20$ (right) for $\omega^2>0$. \\
The energy decreases very slowly for $\bar y_0 \ll 1$ and increases for
large $\bar y_0$. Therefore, there exists a unique minimum of
$V_{eff}(\bar y_0)$ for positive $\bar y_0$. Since the graphs are nearly
flat for $\bar y_0\ll 1$, $y_c$ increases very fast for small
$\lambda_r$ as depicted in Fig. 3. } \label{fig2}
\end{figure}

We thus have the effective potential as a function of $\bar y_0$ by
inserting the solution~(\ref{Om:sol}) to Eq.~(\ref{H:exp2}):
\begin{eqnarray} \label{E:1}
\frac{2V_{eff} (\bar y_0)}{\hbar |\omega|} &=&
\bar\Omega(\bar
    y_0)T(\bar y_0) -
    \frac{\lambda_r}{4\bar\Omega^2(\bar y_0)}
        R(\bar y_0) +\frac{\theta(-\omega^2)}{4\lambda_r}.
\end{eqnarray}
The functional dependence of $V_{eff}(\bar y_0)$ on $\bar
y_0$ is shown in Fig.~\ref{fig2} for a couple of values of
$\lambda_r$.

We write the value $\bar y_0$ that minimizes the effective
potential~(\ref{E:1}) as $y_c$:
\begin{eqnarray} \label{yc}
\left.\frac{d V_{eff}(\bar y_0)}{d\bar y_0}\right|_{\bar
y_0=y_c} =0 .
\end{eqnarray}
Then, the ground state energy is given by $V_{eff}(y_c)$.
With this $y_c$, $\bar \Omega(y_c)$ is determined from
Eq.~(\ref{Om:sol}), and $\epsilon(y_c)$ and $\varepsilon(y_c)$ are
determined from Eqs.~(\ref{vep:bars}) and (\ref{ep:bars}).
With these results the annihilation operator~(\ref{a0:def}) is uniquely
determined.
In Appendix A, we explicitly analyze the variational equation,
Eq.~(\ref{yc}) and the results for positive $\omega^2$ are shown in Fig.
3.

\begin{figure}[htbp]
\begin{center}
\begin{tabular}{cc}
\includegraphics[width=.4\linewidth,origin=tl]{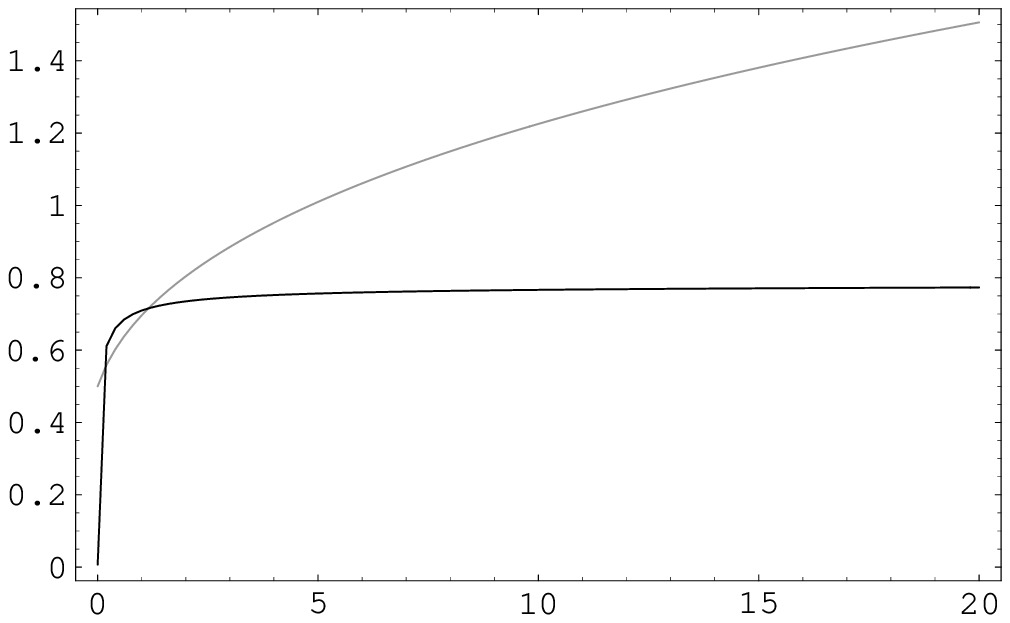} &
\includegraphics[width=.4\linewidth,origin=tl]{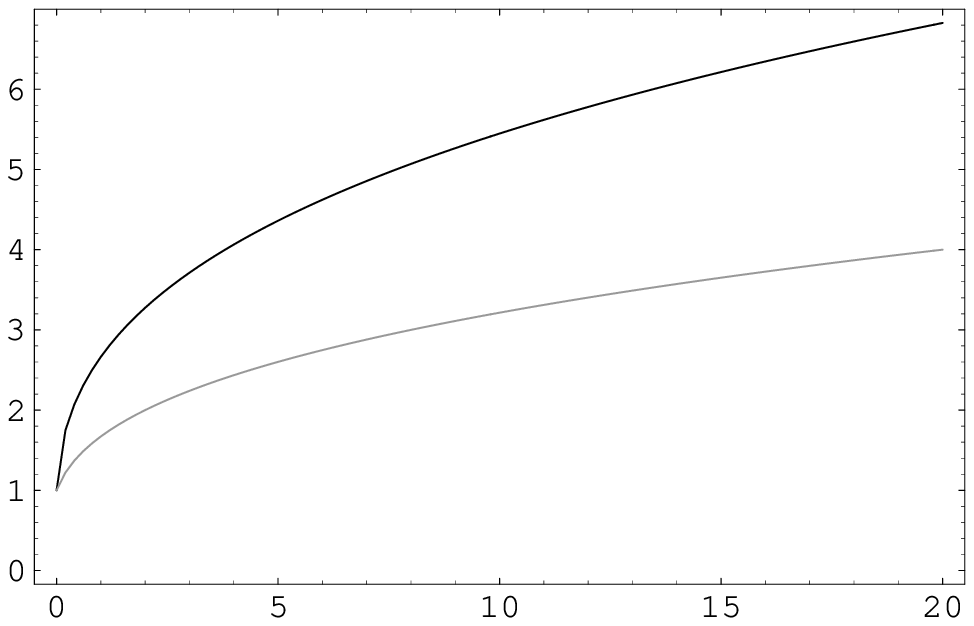}\\
\end{tabular}
\end{center}
\caption{Behavior of ground state energy ($\dsp
\frac{2V_{eff}(y_c)}{\hbar |\omega|}$), $y_c$, $\bar \Omega$, and
$\Omega_G$ as functions of $\lambda_r$
(horizontal axis) for $\omega^2>0$.\\
The left figure shows the energy (grey curve) and $y_c$ (black curve) as
functions of $\lambda_r$.
On the right shows the behavior of the present $\bar \Omega(\lambda_r)$
(black curve) compared to $\Omega_G(\lambda_r)$ (grey curve), the
Gaussian value. As seen in this figure, the effective frequency $\bar
\Omega$ is much larger than the Gaussian value $\Omega_G$ for $\lambda_r
\gg 1$.} \label{fig3}
\end{figure}

In general we have nonzero $y_c$ for positive $\lambda$.
We present the energy eigenvalues of the even and odd ground states and
$\varepsilon$ for several values of the coupling strength in Table 1.
As one can see in Table 1, even for a very small coupling, such as
$\lambda_r \sim 2 \cdot 10^{-3}$, we have non-negligible $y_c$ value,
$\displaystyle \varepsilon= y_c^2/\bar y^2 \sim 0.1$.
Since we have $y_c=0$ in the Gaussian approximation, non-vanishing $y_c$
measures the departure from the Gaussian results.
This is measured by the value $\varepsilon$ in Table 1.
The energy eigenvalues of our zeroth order approximation is closer to
the exact values than the 2nd order perturbative results of the
approximation methods based on the Gaussian approximation for vast range
of couplings.

\begin{table}[htbp]
\caption{The explicit energy eigenvalues for anharmonic and
double-well oscillators for several values of coupling
strength. `Caswell' and `NGAS' refer to the results of
Ref.~\cite{ref7} and \cite{mahapatra}, respectively.
`Exact' implies the exact numerical result.
 }
 \begin{center}
\begin{tabular}{|c|c|c|c|c|c|c||c|c|c|c|}
  \hline & \multicolumn{6}{c||}{$\omega^2 > 0$} &
     \multicolumn{4}{c|}{$\omega^2 < 0$}  \\
 \hline
  $\lambda \hbar/(2|\omega|^3)$ & 0.002 & 0.02 & 0.2 & 2 & 20
    &200 & 200 &20 & 2& 0.2 \\
 \hline
  $\varepsilon $& 0.1054 & 0.2824 & 0.5170 & 0.6419 & 0.6751 & 0.6824
    & 0.6865 & 0.6940 & 0.7296 & 0.8705 \\
 \hline
  $E_0/(\hbar \omega)$ & 0.5007 & 0.5072 & 0.5592 & 0.8041 & 1.5062
    & 3.134 &3.0730 &1.3793 &0.5784& 0.4767 \\
 \hline
  Exact & 0.5007 & 0.5072 & 0.5591 & 0.8038 & 1.5050
    & 3.1314 & 3.0700 & 1.3800 & 0.5800 & 0.4702 \\
 \hline
  NGAS(2nd) & 0.5007 & 0.5072 & 0.5591 & 0.8032 & 1.5030
    & 3.1266 & 3.0650 &1.3752 &0.5752 &0.4606 \\
 \hline
 \hline
    $E_1/(\hbar \omega)$ & ~ & 1.5356 & 1.7696 & 2.7387 & 5.3240
    & 11.193 &11.0393 &4.9984 & 2.0850 & 0.7713 \\
 \hline
  Caswell & ~ & ~ & 1.7695 & 2.7379 &  5.3216
    & 11.187 & 11.002 & 5.0900 & 2.1800 & 0.7703 \\
 \hline
  NGAS(2nd) & ~ & ~ & 1.7694 & 2.7367 & 5.3177 & 11.178
    & 11.024 & 4.9910 & 2.0800 &0.7553  \\
 \hline
\end{tabular}
\end{center}
 \label{table1}
\end{table}

Finally, we consider when the wavefunction becomes aware of the
transition of the potential from a single-well system to a double-well
system in the sense that the wavefunction becomes concave from convex at
$y=0$~\cite{turbiner}. The second derivative of the ground state
wavefunction at $y=0$ is given by
\begin{eqnarray} \label{ddphi}
\left. \frac{\Psi_{0,E}''(y)}{\Psi_{0,E}(y)}\right|_{y=0}=
\nu^2(\nu^2y_0^2-1),
\end{eqnarray}
which changes from negative to positive at $\nu^2 y_0^2=1$.
This equation combined with Eq.~(\ref{vep:ep}) determines
the moment when a the second derivatives of the
wavefunction at $y=0$ changes its sign:
\begin{eqnarray} \label{ssb}
\varepsilon= \frac{2(e+1)}{3e+1} \simeq 0.8123, ~~
\lambda_r\simeq 0.418 .
\end{eqnarray}
This value of critical coupling is somewhat larger than the
corresponding value (0.302) of Ref.~\cite{turbiner} in which the
approximate eigenstates of double-well oscillator were analyzed by
variational approximation with 5-variational parameters.

\subsection{Weak coupling approximations} \label{sub1}
In this subsection, we study the weak coupling ($\lambda_r \ll 1$) limit
of our result.
We need the weak coupling limit for two reasons. First, the approximate
energy~(\ref{H:exp2}) is too complex.
To compare the result with others, it is convenient to consider various
limiting cases.
Secondly, to calculate higher order invariant operators, we need to
understand the behaviors of the Hamiltonian~(\ref{H1:2}) in various
limits of our parameters.
Explicitly, we need to know the orders of magnitudes of the coefficients
$\dsp \left(\frac{\omega^2}{\Omega^2}+\frac{\lambda
    x_0^2}{\Omega^2}\right)\frac{1}{\epsilon}$ and
$ \dsp\frac{\lambda \hbar}{2\Omega^3 \epsilon^2}$ of the quadratic and
quartic parts, respectively, of the Hamiltonian~(\ref{H1:2}) to
construct higher order correction terms.

Before explicitly presenting the limiting case, let us first consider
the Gaussian approximation limit ($y_0 \rightarrow 0$) of our result.
The ground state wavefunction, in this case, is a Gaussian.
For this wavefunction, we have the root mean square value of the
position operator, $\dsp \bar x=x_\Omega=\sqrt{\frac{\hbar}{2\Omega}}
~(\sqrt{3} x_\Omega)$, for the ground state (first excited state).
Therefore, we have $T_E=1$, $\epsilon_{0,E}=1$, and $R_{E}=3$ for even
parity case, and $T_O=3$, $\epsilon_{0,O}=1/\sqrt{3}$, and $R_{O}=5$ for
odd parity case.
The value of $\Omega$ is determined by the Gap equation:
\begin{eqnarray*}
\left(\Omega_r\right)^3-\Omega_r
    -\lambda_r R =0 .
\end{eqnarray*}
This gap equation is the same as that of the Gaussian approximation for
the state $|0_E\rangle$.
The gap equation for the state $|0_O\rangle$ is different from that of
the Gaussian approximation since $R_O\neq R_E$.
This difference makes the energy of the $1^{st}$ excited state different
from that of the Gaussian approximation.
The energies for both states are
\begin{eqnarray*}
E_{0,E/O}= \frac{\hbar \Omega_{E/O}
T_{E/O}}{2}\left(1-\frac{\lambda \hbar}{8 \Omega_{E/O}^3}
R_{E/O} \right) .
\end{eqnarray*}
The explicit value of $E_{0,O}$ is slightly larger than that of the
Gaussian approximation and closer to the exact value of the anharmonic
oscillator.
This result coincides with the numerical energy eigenvalues of
Ref.~\cite{mahapatra} for quantum anharmonic oscillators, in which a New
Gaussian Approximation Scheme (NGAS) was used.
%

We now consider the small coupling limit of our result for
positive frequency squared ($\lambda_r \ll 1$,
$\omega^2>0$) case. In this case, the solution of the gap
equation~(\ref{dOm=0}) becomes
\begin{eqnarray} \label{gap2}
\bar\Omega(\bar y_0) = \bar \Omega_0\left[1+\frac{\lambda_r
R}{2\Omega_0^3 T}-\frac{3}{2}\left(\frac{\lambda_r R}{2\bar
\Omega_0^3 T}\right)^2 +O(\lambda_r^3)\right] ;~~~~ \bar
\Omega_0^2 = 1+\frac{2\bar y_0^2}{T} .
\end{eqnarray}
In addition, the variational equation~(\ref{yc}) becomes
\begin{eqnarray} \label{yc2}
\frac{d}{d\bar y_0}\left[T^{1/2}\sqrt{T+2\bar
y_0^2}\left(1+\frac{\lambda_r R }{4\bar \Omega_0^3 T}-
    \left(\frac{\lambda_r R}{2\bar \Omega_0^2
    T}\right)^2
    +O(\lambda_r^3)\right)\right]_{\bar y_0=y_c} =0.
\end{eqnarray}
The $O(\lambda_r^0)$ solution to Eq.~(\ref{yc2}) is $\bar y_0=0$.
Therefore, we can series expand Eq.~(\ref{yc2}) around $\bar y_0=0$:
\begin{eqnarray*} \label{yc3}
\frac{d}{d\bar
y_0}\left[1+\frac{3\lambda}{2}+\frac{9\lambda_r^2}{8}
    -\frac{\lambda\bar  y_0^4}{2}+
    \frac{\bar y_0^8}{12}-\frac{\bar y_0^{12}}{60}+O(\bar
    y_0^{16})
    \right]\simeq 2\bar y_0^3\left(-\lambda
    +\frac{\bar y_0^4}{3}\right)+\cdots=0 .
\end{eqnarray*}
Thus, the minimizing value $y_c$ of the effective potential and $\bar
\Omega$ become
\begin{eqnarray} \label{etc:s}
y_c &=& (3\lambda_r)^{1/4}, \quad \bar \Omega(y_c)=1+\sqrt{3\lambda_r}+
3\lambda_r+O(\lambda_r^{3/2}).
\end{eqnarray}
With this the energy becomes
\begin{eqnarray}
\frac{2V_{eff}}{\hbar \omega} &=& 1+
\frac{3\lambda_r}{2}+\frac{3\lambda_r^2}{8}+O(\lambda_r^3).
\nn
\end{eqnarray}
From Eqs.~(\ref{vep:bars}) and (\ref{ep:bars}), we have
\begin{eqnarray} \label{..}
\varepsilon = 2 y_c^2-2 y_c^4 +\cdots, \quad\quad \epsilon
= 1-y_c^2+2 y_c^4 +\cdots .
\end{eqnarray}
Therefore, we find that the correction term in $V_{eff}$ is of
$O(\varepsilon^2)\sim O(\delta^2)$ and the correction term in $\bar
\Omega$ is of $O(\varepsilon)\sim O(\delta)$.

We next consider the $\lambda_r \ll 1$ limit with negative frequency
squared case, $\omega^2<0$.
In the limit $y_0\rightarrow \infty$, we have the exact result
$\bar\Omega = \sqrt{2}$ and $\bar y_0^2 = (\sqrt{2} \lambda_r)^{-1}$ for
$\lambda_r \rightarrow 0$.
Therefore, to have series expansion around the limit, we use the
following change of variable from $\bar y_0$ to $z$:
\begin{eqnarray*}
 \bar y_0^2=\frac{1}{\sqrt{2}\lambda_r(1+z)}.
\end{eqnarray*}
We have  $z=0$ in the limit $\lambda_r\sim 1/(\sqrt{2} \bar y_0^2)
\rightarrow 0$. Now, the solution of the gap equation~(\ref{dOm=0})
becomes
\begin{eqnarray} \label{gap:3}
\bar \Omega &=& \bar \Omega_0\left[1-(1+\bar
    \Omega_0^2)\frac{\lambda_r T(1+z)}{\sqrt{2}}
+(1+\bar \Omega_0^2)(1+3\bar \Omega_0^2)
    \left(\frac{\lambda_r T(1+z)}{\sqrt{2}}
    \right)^2+O(\lambda_r^3)\right]; \\
\bar \Omega_0&=&\frac{\sqrt{2} R}{4\bar y_0^4(1+z)} ,\nn
\end{eqnarray}
where $T$ and $R$ are implicitly assumed to be functions of
$z$:
\begin{eqnarray*}
T= 1+O(e^{-(\sqrt{2}\lambda_r)^{-1}}) ,
 \quad \frac{1}{\lambda_r}\left(
    \frac{R}{4\bar y_0^4}-1\right)
    =3\sqrt{2}(1+z)+
    \frac{3\lambda_r}{2}(1+z)^2
    +O(e^{-(\sqrt{2}\lambda_r)^{-1}}) .
\end{eqnarray*}
For small $\lambda_r$, we essentially set $T=1$.
The effective energy~(\ref{E:1}) now becomes
\begin{eqnarray} \label{E:dw}
\frac{2V_{eff} (z)}{\hbar |\omega|} &=& \bar\Omega(\bar
    y_0)T(\bar y_0) -\frac{\bar \Omega_0^2}
        {4\bar \Omega^2\lambda_r}\left(\frac{4\bar
    y_0^4}{R} -1\right)-
    \frac{1}{4\lambda_r}\left(\frac{\bar \Omega_0^2}
        {\bar \Omega^2} -1\right) \\
&=& \Omega_0+\frac{(2-\Omega_0^2)(1+z)}{2\sqrt{2}}
    +\frac{(1+z)\lambda_r}{8}\left[-4\sqrt{2}\Omega_0(
    \Omega_0^2+1)+(3\Omega_0^4+14\Omega_0^2-22)(1+z)
    \right]+O(\lambda_r^2) \nn .
\end{eqnarray}
For small $\lambda_r$,
\begin{eqnarray} \label{dE/dz}
\frac{2}{\hbar |\omega|}\frac{d V_{eff} (z)}{dz} &=&
\sqrt{2} z-\frac{9}{2}\lambda_r +O(\lambda_r^2) .
\end{eqnarray}
Therefore, we have $\displaystyle z=\frac{9\lambda_r}{2\sqrt{2}}$ and
$\dsp \bar\Omega=\sqrt{2}-\frac{3\lambda_r}{2}+\cdots $ in the small
coupling limit.
With this value, we have $\displaystyle \frac{2E (z)}{\hbar
|\omega|}=\sqrt{2}-\frac{3\lambda_r}{4} +O(\lambda_r^2)$.
The explicit dependence of $\varepsilon$ and $\epsilon$ on $\lambda_r$
can be obtained from Eqs.~(\ref{ep:bars}) and (\ref{vep:bars}):
\begin{eqnarray} \label{ep:3}
\varepsilon = 1-\frac{\lambda_r}{\sqrt{2}}+O(\lambda_r^2),
\quad \quad  \epsilon =
\frac{\sqrt{\lambda_r}}{2^{1/4}}+O(\lambda_r^{3/2}) .
\end{eqnarray}
We thus find that, the correction to the energy starts from the terms of
$O(\epsilon^2)\sim O(\delta^2)$.

With these results for the small coupling cases, we conclude that the
correction to the energy starts from the order $\delta^2$.
In addition, the corrections to the frequency and other parameters
starts from the order $\delta$.

\section{Higher order construction of the creation and annihilation
operators}
In this section, we develop a general procedure to generate perturbative
series of the invariant creation and the annihilation operators based on
the $0^{th}$ order invariant operators obtained in the previous section.
First, we explicitly calculate the $1^{st}$ order creation and
annihilation operators from the $0^{th}$ order operators.
Then, we write a general formula which gives the $n^{th}$ order
invariant creation and annihilation operators from $(n-1)^{th}$ order
ones.
Some formulas for these calculations are given in Appendix B, C, and D.

We first consider the structure of the Hamiltonian~(\ref{H1:2}) which
can be written in the form,
\begin{eqnarray} \label{H:31}
H \equiv \frac{\hbar \Omega}{2}\cH +E_0,
\end{eqnarray}
where $E_0$ is the ground state energy and the operator
$\cH$ represents the operator part of the Hamiltonian with
vanishing ground state expectation value,
\begin{eqnarray} \label{cN1}
\cH =\frac{1}{2}T_2^0+\frac{\epsilon h_1}{\epsilon_0^2}
T_0^2+ \frac{\varepsilon
    h_2}{2}T_0^4-n_0;\quad\quad
    n_0 = \langle 0|\left(\frac{1}{2}T_2^0+\epsilon h_1
    T_0^2+ \frac{\varepsilon h_2}{2}
    T_0^4\right)|0\rangle .
\end{eqnarray}
Following the principle of variational perturbation theory, we use for
the parameters $\Omega$ and $y_0$ in Eqs.~(\ref{H:31}) and (\ref{cN1})
their zeroth order values determined in the previous section.
Note that this form~(\ref{cN1}) resembles $H_m$ in Eq.~(\ref{H:sim})
except for small differences in the coefficients $h_i$.
The zeroth order Hamiltonian $H_m$ is obtained by setting $h_i=1$ in
$\cH$.
In the Appendix B, we show that the coefficients $h_i$ are given, in
fact, by $O(\delta^0)$ numbers by using the zeroth order result.
The quantities
\begin{eqnarray} \label{barh1}
\bar h_1&=& \frac{\epsilon(h_1-1)}{\delta\epsilon_0^2},
 \quad \quad \bar h_2=\frac{\varepsilon(h_2-1)}{2\delta}
  ,
\end{eqnarray}
are also $O(\delta^0)$ numbers.
In addition, it is shown that, $\epsilon \bar h_1$ and $\varepsilon \bar
h_2$ are of $O(\delta)$ in the Appendix B.
With the equation~(\ref{barh1}), the Hamiltonian~(\ref{H:31}) can be
written as the sum of $H_0$ of (\ref{H:sim}) and the correction terms:
\begin{eqnarray} \label{H:32}
H &=& \frac{\hbar \Omega}{2}\left(H_m + \delta \Delta
H\right)+E_0-\frac{\hbar \Omega}{2}n_0,
\end{eqnarray}
where the perturbation term $\Delta H$ is
\begin{eqnarray} \label{H-H1}
\Delta H =(\cH +n_0- H_m)/\delta = \bar h_1 T_0^2+ \bar
h_2T_0^4 \,.\nn
\end{eqnarray}

We now show that every even power polynomials of $\hat p$ and $\hat y$
can be written in terms of the invariant operators.
From Eqs.~(\ref{cN1}) and (\ref{ab:sol0}) we write $T_m^n~(m+n=1)$ as a
function of $\ca$, $\ca^\dagger$, and $ \cH$:
\begin{eqnarray} \label{T:a}
T_1^1 &=& \frac{1}{2}\left(\ca
    +\ca^\dagger\right),~~~\\
T_0^2 &=& \frac{\bar \alpha^2}{2 \delta
h_2}\left(-1+\sqrt{1+
    \frac{4\delta h_2}{\bar \alpha^4}
    \hat B} \right);\quad
\bar \alpha^2 =\alpha \nu^2+ \frac{2h_1\epsilon^2}{
     \epsilon_0^2}, \nn\\
T_2^0 &=& 2\left(\cH+n_0\right)- \frac{2\epsilon
h_1}{\epsilon_0^2} T_0^2- \varepsilon h_2 (T_0^2)^2 , \nn
\end{eqnarray}
where $\hat B(t)$ is an $O(\delta^0)$ operator,
\begin{eqnarray} \label{B}
\hat B(t)=2\epsilon
    (\cH+n_0)- \frac{\alpha}{2i}\left(\ca-
\ca^\dagger \right) .
\end{eqnarray}
Note also that the difference between the coefficients $\bar
\alpha^2-\alpha^2$ is
\begin{eqnarray} \label{diff:alpha}
\bar\alpha^2-\alpha^2
    =2\delta \epsilon \bar h_1 \sim O(\delta^2) .
\end{eqnarray}
Expanding $\hat T_0^2$ in series of $\delta$, we get a well defined
series expansion as a function of $\hat B$,
\begin{eqnarray} \label{T02:delta}
\hat T_0^2 = \frac{\hat B}{\bar
\alpha^2}\left[1-\frac{\delta h_2}{\bar\alpha^4}\hat B +
2\left(\frac{\delta h_2}{\bar\alpha^4}\hat B\right)^2
-5\left(\frac{\delta h_2}{\bar\alpha^4}\hat
B\right)^3\right]+ O(\delta^4); \quad\quad
 |\hat B| < \frac{\bar \alpha^4}{4\delta h_2} ,
\end{eqnarray}
where $|\hat B|$ implies the ground state expectation value of the
operator $\hat B$.
Since each even power polynomial of $\hat \pi$ and $\hat y$ can be
written as a sum of the multiplications of $T_2^0$, $T_1^1$, and
$T_0^2$, they can also be written in terms of $\cA$, $\cA^\dagger$, and
$\cH$.

We now develop the procedure to obtain the higher order invariant
operators from the zeroth order ones.
The invariant operators satisfy the LvN equation,
\begin{eqnarray} \label{LvN:1st2}
i \hbar \frac{d \hat A(\hat p(t),\hat x(t),t)}{dt} = i
\hbar \frac{\partial
    \hat A}{\partial t}+[\hat A,H_m]
    +\delta[\hat A, \Delta H]=0.
\end{eqnarray}
where it should be noted that $H_m$ and $\Delta H$ are
written as a functions of $\cA$, $\cA^\dagger$, and $\cH$.
For example,
\begin{eqnarray*} \label{H;DH}
\Delta H&=& \frac{\bar \alpha^2 \bar h_1}{2\delta h_2}
    \left(-1 +\sqrt{1+\frac{4\delta h_2}{\bar \alpha^4}
   \hat B}\right)+\frac{\bar \alpha^4 \bar h_2}{4\delta^2 h_2^2}
    \left(-1 +\sqrt{1+\frac{4\delta h_2}{\bar \alpha^4}
    \hat B}\right)^2 \\
    &\simeq & \frac{\bar h_1}{\bar \alpha^2} \hat B+
        \frac{\bar h_2}{\bar \alpha^4} \hat B^2+\cdots .
        \nn
\end{eqnarray*}
One may search for the higher order invariant operators in the form:
\begin{eqnarray} \label{A:gen}
\hat A =f(\alpha_i(t), \cA,\cA^\dagger,\cH) ,
\end{eqnarray}
where $\alpha_i$'s denote the coefficient functions of the polynomials
of $\cA$, $\cA^\dagger$, and $\cH$.

Since $\cA$, $\cA^\dagger$, and $\cH$ satisfy the LvN
equation to $O(\delta^0)$, any function $f(\alpha_i(t),
\cA,\cA^\dagger,\cH)$ of these operators satisfies
\begin{eqnarray} \label{..LvN3}
i\hbar \frac{d f(\alpha_i(t), \cA,\cA^\dagger,\cH)}{dt}&=&
i\hbar \frac{\partial f(\alpha_i(t),
\cA,\cA^\dagger,\cH)}{\partial t}+[f(\alpha_i(t),
\cA,\cA^\dagger,\cH), H]\\
&\simeq&\sum_i \dot \alpha_i(t) \frac{\partial
f(\alpha_i(t), \cA,\cA^\dagger,\cH)}{\partial \alpha_i}
.\nn
\end{eqnarray}
This equation converts the operator equation, the LvN equation, into a
differential equation for the coefficients functions $\alpha_i$.

\subsection{Construction of the first order
invariant operators}
In this subsection, we construct the first order invariant
annihilation operator from the zeroth order invariants.
The LvN equation is solved to the first order in $\delta$ and the
algebra satisfied by the invariant operators are obtained.
The $1^{st}$ order ground state and Hilbert space are constructed based
on these first order invariant operators.

%

Since the zeroth order invariant annihilation operator is
$\cA$, the first order invariant annihilation operator,
which satisfies the LvN equation~(\ref{LvN}), should be
given by adding $O(\delta)$ correction term.
Considering nontrivial accumulation of phase factor, we write the first
order invariant annihilation operator as in the following ansatz:
\begin{eqnarray} \label{gen1}
\hat A_1 &=&e^{i \hat \Psi_1(t)- i \alpha \Omega t}
   \left( \cA+\delta
    \hat A_{(1)}\right),
\end{eqnarray}
where the phase operator, $\hat \Psi_1(t)$, and the
$1^{th}$ order correction term, $\hat A_{(1)}$, are
\begin{eqnarray}\label{corr1}
\hat \Psi_1(t) &=& \Psi(0) + \Omega \left[\alpha t +
    \delta \int^t dt'
    {\cal H}_1(t')\right];~~~~ {\cal H}_1=
        h_{10}(t)+ h_{11}(t)\cH ,  \\
\hat A_{(1)}&\equiv& b_{(1)}^{0,0}+
    b_{(1)}^{1,0} \cA^\dagger +
   b_{(1)}^{0,1} \cH+b_{(1)}^{2,0} \hat
   {\cal A}^{\dagger 2} + b_{(1)}^{0,2} \cH^2
   +b_{(1)}^{1,1} \cA^\dagger \cH+
    c_{(1)}^{0,2} \cA^2 . \nn
\end{eqnarray}
The operators in $\hat A_{(1)}$ are ordered as in the normal ordering.
$\Psi(0)$ is an arbitrary real constant number.
A nontrivial operator, $\cH$, in the phase operator $\hat \Psi_1(t)$ is
included to remove the unphysical linearly increasing term which was
present in doing similar calculation for anharmonic oscillator in
Ref.~\cite{bak}.
The terms of the form $\cA$ and $\cH \cA$ in $\hat A_1$ are generated by
the phase operator.
Therefore, such terms are omitted in $\hat A_{(1)}$.

Now, we substitute the operator $\hat A_1$ to the LvN
equation~(\ref{LvN}).
With the help of Eq.~(\ref{..LvN3}), we show that, to the first order in
$\delta$, the LvN equation can be solved by a simple comparison of
coefficients functions with the ansatz~(\ref{gen1}).
To the first order in $\delta$, the invariant annihilation operator
satisfies,
\begin{eqnarray} \label{LvN:1st}
0\cong\frac{i \hbar}{e^{i\hat \Psi(t)-i\alpha\Omega t}}
\frac{d \hat A_1(t,\hat p(t),\hat x(t))}{dt}
 &=&-\hbar \delta \Omega {\cal H}_1(\cA+ \delta \hat A_{(1)})
    +i\hbar \frac{d}{dt}\left( \cA+\delta
    \hat A_{(1)}\right) \\
&=&i \hbar \frac{d
    \cA}{d t}+\hbar \delta \left[
    i\frac{d \hat A_{(1)}}{dt}-\Omega
        {\cal H}_1(\cA+ \delta \hat
A_{(1)})\right], \nn
\end{eqnarray}
where $\cong$ implies ``the same up to $O(\delta^1)$".
Note that the zeroth order invariant operators $\cA$ and $\cA^\dagger$
in Eq.~(\ref{LvN:1st}) are given by $\cA=b_0e^{i\alpha \Omega t}
\ca(\hat p(t),\hat x(t))$ and $\cA^\dagger=b_0e^{-i\alpha \Omega t}
\ca^\dagger(\hat p(t),\hat x(t))$ as in (\ref{ab:sol0}).
To the zeroth order in $\delta$, the total derivative $d\cA/dt $ of the
invariant operator vanishes.
Therefore, $d\cA/dt$ is an $O(\delta^1)$ operator.
Thus Eq.~(\ref{LvN:1st}) reduces to:
\begin{eqnarray*}
i \frac{d\hat A_{(1)}}{dt}\simeq \Omega {\cal H}_1 \cA -
\frac{i}{\delta} \frac{d\cA}{dt}.
\end{eqnarray*}
By using the definition of the total derivative, the LvN equation can be
written as
\begin{eqnarray} \label{1st:0}
i\frac{d \hat A_{(1)}}{dt} &\simeq &
    \Omega {\cal H}_1\cA-\frac{1}{\delta\hbar} \left(i \hbar
    \frac{\partial \cA}{\partial t}+
    [\cA, H]\right) \\
 &=& \Omega\left[ {\cal H}_1\cA+\frac{1}{2\delta} \left(2 \alpha
    \cA- [\cA, \cH]\right)\right] . \nn
\end{eqnarray}
To the zeroth order in $\delta$, the left hand side of this equation,
the total derivative $\displaystyle \frac{d \hat A_{(1)}}{dt}=
\frac{\partial\hat A_{(1)}}{\partial t}+ \frac{[\hat
A_{(1)},H]}{i\hbar}$, is given by simply taking the derivatives of the
coefficients, because $\displaystyle \frac{d\cA}{dt}\simeq 0$ and
$\displaystyle \frac{d \cH}{dt}\simeq 0$.
Therefore, we have
\begin{eqnarray} \label{1st:2}
\frac{dA_{(1)}}{dt}&\simeq & \dot b_{(1)}^{0,0}+ \dot
    b_{(1)}^{1,0} \cA^\dagger +
   \dot b_{(1)}^{0,1} \cH+\dot b_{(1)}^{2,0} \hat
   {\cal A}^{\dagger 2} + \dot b_{(1)}^{0,2} \cH^2
   +\dot b_{(1)}^{1,1} \cA^\dagger \cH+
   \dot c_{(1)}^{0,2} \cA^2 .
\end{eqnarray}
The difference $2\alpha \cA-[\cA, \cH]$ in the right hand side of
Eq.~(\ref{1st:0}) is computed in Appendix B:
\begin{eqnarray*} \label{diff222}
\frac{2\alpha \cA-[\cA, \cH]}{
    \delta b_0 e^{i\alpha \Omega t}}
&=& \frac{2ih_1}{\epsilon_0^2}+4i(\epsilon \bar
    h_1+\varepsilon \bar
    h_2)T_0^2-\frac{4\epsilon \bar h_1}{\alpha}
    T_1^1-2h_2\left(-2i T_0^4+
    \frac{\{T_0^2, T_1^1\}}{\alpha}\right) \\
&\simeq & \frac{2ih_1}{\epsilon_0^2}-\frac{2h_2}{\bar
\alpha^2}\left(-\frac{2i}{\bar \alpha^2}
    \hat B^2+ \frac{\{\hat B, \ca
    +\ca^\dagger\}}{2\alpha }\right), \nn
\end{eqnarray*}
where we have used the fact that both of the number
$\epsilon \bar h_1$ and $\varepsilon \bar h_2$ are of
$O(\delta)$.
Using the explicit formula for $\hat B$, $\hat B^2$, and $\{\hat B,
\ca\}_+$ in Eqs.~(\ref{B}), (\ref{b2}), and (\ref{A0:Hm}) we have
\begin{eqnarray} \label{diff222}
\frac{2\alpha \cA-[\cA, \cH]}{
    \delta b_0 e^{i\alpha \Omega t}}
&\simeq & i \,h(\cH)- \frac{ih_2}{\bar \alpha^2}\left[
 \left(\frac{\alpha^2}{\bar \alpha^2}+1\right)\ca^2+
 \left(\frac{\alpha^2}{\bar \alpha^2}-1\right)\ca^{\dagger 2}
 \right]  \\
&-&\frac{4\epsilon h_2}{\alpha \bar
\alpha^2}\left[\left(\frac{2
    \alpha^2}{\bar \alpha^2}+1\right)(\cH+n_0+\alpha)\ca -
    \left(\frac{2
    \alpha^2}{\bar \alpha^2}-1\right)\ca^\dagger
    (\cH+n_0+\alpha)\right],
\nn
\end{eqnarray}
where
\begin{eqnarray} \label{hfn}
h(\cH) = \frac{2h_1}{\epsilon_0^2}
    +\frac{4h_2}{\bar \alpha^2}\left[
    \frac{3\epsilon^2}{2}+\varepsilon(\cH+
    n_0)+ \frac{6\epsilon^2}{\alpha^2}(\cH+n_0)^2\right].
\end{eqnarray}
Therefore, with Eqs.~(\ref{1st:2}) and (\ref{diff222}), the LvN
equation~(\ref{1st:0}) becomes
\begin{eqnarray} \label{LvN:final}
\dot b_{(1)}^{0,0}&+& \dot
    b_{(1)}^{1,0} \cA^\dagger +
   \dot b_{(1)}^{0,1} \cH+\dot b_{(1)}^{2,0} \hat
   {\cal A}^{\dagger 2} + \dot b_{(1)}^{0,2} \cH^2
   +\dot b_{(1)}^{1,1} \cA^\dagger \cH+
   \dot c_{(1)}^{0,2} \cA^2 \\
&=& \frac{\Omega}{2i}\left\{ 2h_{10}\cA+2h_{11}\cH \cA
 +
 i\,h(\cH) b_0e^{i\alpha \Omega t}
    \nn \right.\\
&-&\left. \frac{2ih_2}{b_0 \alpha^2}
    e^{-i\alpha\Omega t}\cA^2
- \frac{4\epsilon h_2}{
    \alpha^3}\left[3(\cH+n_0+\alpha)\cA
    - e^{2i\alpha \Omega t} \cA^\dagger
    (\cH+n_0+\alpha)\right]
    \right\} . \nn
\end{eqnarray}
Now the LvN equation has become a simple relation between
the coefficients of Eq.~(\ref{LvN:final}). By comparing the
coefficients of $\cA$ and $\cH \cA$, we get
\begin{eqnarray} \label{1st:sol}
h_{10} &=& \frac{6 \epsilon (\alpha +n_0)h_2}{\alpha^3},
 \quad\quad h_{11}= \frac{6\epsilon h_2}{\alpha^3}.
\end{eqnarray}
After inserting Eq.~(\ref{1st:sol}) into Eq.~(\ref{LvN:final}), we
obtain  $\hat A_{(1)}$ by simply integrating the coefficients on the
right hand side of the equation~(\ref{LvN:final}).
Formally, we may write the formula as:
\begin{eqnarray} \label{A1}
A_{(1)}(t)\simeq \frac{\Omega}{2i}
  \int^t dt\left\{2{\cal H}_1 \cA
    +\frac{2\alpha
    \cA-[\cA, H_m]}{\delta}\right\}.
\end{eqnarray}
In summary, the creation operator and the annihilation
operator to $O(\delta)$ are
\begin{eqnarray} \label{ops:1st}
\hat A_1^\dagger= b\hat a_1^\dagger e^{-i \hat \Psi_1(t)}
=b ( \ca^\dagger + \hat a_{(1)}^\dagger)e^{-i \hat
    \Psi_1(t)}, ~~~~
 \hat A_1=be^{i \hat \Psi_1(t)} \hat a_1
    = be^{i \hat \Psi_1(t)}
 ( \ca + \hat a_{(1)}),
\end{eqnarray}
where the phase operator is
\begin{eqnarray} \label{phase:t}
\hat \Psi_1(t) = \Psi(0)+\alpha \Omega t\left[1+
 \frac{6\epsilon \delta h_2}{\alpha^4}
 \left(\cH+n_0+\alpha\right)\right] .
\end{eqnarray}
The first order correction terms in Eq.~(\ref{ops:1st}) are
\begin{eqnarray} \label{A1;1}
\hat a_{(1)}\equiv \frac{\hat A_{(1)}(t)}{be^{i \alpha
\Omega t}} &=&
 -\frac{i}{2\alpha}h(\cH)
    -\frac{\epsilon h_2}{
        \alpha^4}\ca^\dagger \left(\cH+n_0+\alpha\right)-
 \frac{ih_2}{\alpha^3}
    \ca^2, \\
\hat a_{(1)}^\dagger \equiv \frac{\hat
    A_{(1)}^\dagger(t)}{be^{-i \alpha \Omega t}}
&=& \frac{i}{2\alpha}h(\cH)
    -\frac{\epsilon h_2}{
        \alpha^4}\left(\cH+n_0+\alpha \right)\ca +
 \frac{ih_2}{\alpha^3}
    \ca^{\dagger 2}. \nn
\end{eqnarray}

The zeroth order invariant operators satisfy a generalized deformed
algebra~(\ref{ada}). It is an interesting question if this algebraic
structure is preserved to the first order.
To this end, we consider the algebra satisfied by the first order
creation and the annihilation operators, $\hat A_1^\dagger$ and $\hat
A_1$.
The commutator of the first order operators becomes
\begin{eqnarray} \label{A1:comm}
\frac{[\hat A_1, \hat A_1^\dagger]}{4b^2} &\cong & \nu^2
\varepsilon
     +\frac{4\epsilon^2(\cH+n_0)}{\alpha}
      -\frac{3\epsilon^3 \delta h_2}{
        \alpha^3}\left(1+\frac{4(\cH+n_0)^2}{\alpha^2}\right)
        \, .
\end{eqnarray}
%
The anticommutator becomes
\begin{eqnarray}\label{A1:anti}
 \frac{\{\hat A_1,\hat A_1^\dagger\}_+ }{4b^2}&\cong &
 \frac{3\epsilon^2}{2}+
    \left(\varepsilon-\frac{2h_1 \epsilon \delta}{
        \epsilon_0^2\alpha^2}-\frac{3h_2\epsilon^3
        \delta}{\alpha^4}\right)(\cH+n_0) +
    \frac{2\epsilon^2}{\alpha^2}(\cH +n_0)^2
    -\frac{12 h_2\epsilon^3 \delta}{\alpha^6}
    (\cH+n_0)^3
    ,
\end{eqnarray}
where $b^2 = b_0^2(1 + \delta b_{(1)})$.
The time-dependent part of the phase factor~(\ref{phase:t}) can be
understood by calculating the commutator of $\hat a_1$ and $\cH$,
\begin{eqnarray*} \label{..3}
~[\hat a_1, \cH] \cong 2\alpha \left[1+\frac{6\epsilon
    h_2\delta}{\alpha^4} \left(\cH+n_0+\alpha\right)\right]
    \hat a_1  ,
\end{eqnarray*}
where the form in the square bracket of the right-hand side is
remarkably the same as that appears in the phase factor~(\ref{phase:t}).
Thus the phase operator appears to have an imprint of the energy
difference of two neighboring states in each parity sector.

For these commutator~(\ref{A1:comm}) and anticommutator~(\ref{A1:anti})
to be those of a generalized deformed oscillator, the algebra must be of
the following form:
\begin{eqnarray*} \label{phi:genform}
\hat A_1 \hat A_1^\dagger &= & \Phi_1(\hat N_1+1), ~~~~
    \hat A_1^\dagger\hat A_1 = \Phi_1(\hat N_1),
\end{eqnarray*}
where the structure function satisfies $\Phi_1(0)=0$.
Assuming the operator $\cH$ in Eq.~(\ref{cN1}) can be written in terms
of the first order number operator $\hat N_1$,
\begin{eqnarray*}\label{1st}
\cH(\hat N_1)&=& 2 \alpha \hat N_1+ \delta(E_1+n_1\hat N_1
+n_2\hat N_1^2) ,
\end{eqnarray*}
we can compare the coefficients of $(\hat N_1)^a$ term of $\hat A_1\hat
A_1^\dagger$ with that of $\dsp \left.\hat A_1^\dagger \hat
A_1\right|_{\hat N_1\rightarrow \hat N_1+1}$ for $a=0,1,2,3$ from
Eqs.~(\ref{A1:comm}) and (\ref{A1:anti}).
Rather than directly computing the coefficients of $(\hat N_1)^a$, it is
convenient to compare the coefficients of the powers of $(\cH+n_0)$ in
$\dsp \left.\hat A_1^\dagger \hat A_1\right|_{\hat N_1\rightarrow \hat
N_1+1}$ with those of $\hat A_1\hat A_1^\dagger$ by using the identity:
\begin{eqnarray*}
\cH(\hat N_1+1) +n_0=
2\alpha'+\left(1+\frac{n_2\delta}{\alpha} \right)[\cH(\hat
N_1)+n_0];\quad \quad
\alpha'=\alpha+\frac{\delta}{2}\left(n_1+n_2-\frac{
    n_0n_2}{\alpha} \right)\,,
\end{eqnarray*}
where it is assumed that $\cH(\hat N_1)$ is a function of $\hat N_1$.
The coefficients of $(\cH+n_0)^3$ terms do not provide any information,
and the coefficients of $(\cH+n_0)^2$ and $(\cH+n_0)$ determine $n_1$
and $n_2$:
\begin{eqnarray} \label{n0}
n_1=\frac{12
    \epsilon h_2 n_0}{\alpha^3}, \quad \quad
    n_2=\frac{12\epsilon h_2}{\alpha^2} .
\end{eqnarray}
The zeroth order term in $(\cH+n_0)$ expansion is automatically
satisfied with these $n_i$'s, which is a non-trivial requirement to be
an algebra.
The energy $E_1$ is determined by the condition $\Phi_1(\hat N_1=0)=0$,
which leads to a cubical equation
\begin{eqnarray} \label{E0'}
\frac{\Phi_1(0)}{2b^2}
=\frac{3\epsilon^2}{2}-\nu^2\varepsilon+\frac{3\epsilon^3
    \delta}{\alpha^3}
 +\left(\varepsilon-\frac{4\epsilon^2}
    \alpha -\frac{2 h_1\epsilon\delta}{\epsilon_0^2\alpha^2}-
    \frac{3h_2\epsilon^3\delta}{\alpha^4}\right)E_0'+
    \left(\frac{2\epsilon^2}{\alpha^2}+\frac{12 h_2\epsilon^3
    \delta}{\alpha^5}\right){E_0'}^2-
    \frac{12h_2\epsilon^3\delta}{\alpha^6} {E_0'}^3 \cong
    0,
\end{eqnarray}
where $E_0'=n_0+\delta E_1$. 


Using Eqs.~(\ref{n0}) and (\ref{E0'}) we get the structure function
$\Phi_1(\hat N_1)$ in a series form:
\begin{eqnarray}\label{1st}
 \Phi_1(\hat N_1) &\cong & \Phi_0(\hat N_1)+ \delta(\phi_1
\hat N_1 + \phi_2 \hat N_1^2 +\phi_3 \hat N_1^3),
 \\
\cH(\hat N_1)+n_0&\cong & E_0'+(2 \alpha+\delta n_1) \hat
    N_1+ \delta n_2\hat N_1^2 , \nn
\end{eqnarray}
where $\phi_i$ are determined to be
\begin{eqnarray} \label{phi}
\phi_1=4b^2(\cE_+-2 \bar \epsilon \delta) ,\quad \quad
\phi_2 = 16 b^2 \epsilon^2, \quad \quad \phi_3 =0 .
\end{eqnarray}
Here the coefficients ${\cal E}_+$ and $\bar \epsilon$ are
\begin{eqnarray} \label{bep}
{\cal E}_+=\alpha\varepsilon
    -4\epsilon^2\left(1-\frac{E_0'}{\alpha}\right), \quad
\bar\epsilon=\frac{\epsilon}{\alpha^3}\left(
    \frac{\alpha^2h_1}{\epsilon_0^2}
    +\frac{3h_2\epsilon^2}{2}+
    \frac{6h_2\epsilon^2{E_0'}^2}{\alpha^2}\right).
\end{eqnarray}%
From $\Phi_1(1)=1$, we determine the explicit value of $b^2$,
\begin{eqnarray} \label{b1}
4 b^2 = \frac{1}{{\cal E}_+'-2\bar \epsilon \delta
+4\epsilon^2} .
\end{eqnarray}
Notably, the structure function $\Phi_1(\hat N_1)$ does not
get correction term of the form $\hat N_1^3$ since $\phi_3=
0$, even though the operator $\cH$ gets an additional
correction term of $\hat N_1^2$ due to the $O(\delta)$
calculation.
This implies that the form of the structure function does not change
except for the $O(\delta)$ correction in coefficients.

In summary, the creation operator, $\hat A_1^\dagger$, and the
annihilation operator, $\hat A_1$, satisfy the algebraic relation,
\begin{eqnarray} \label{ada1}
\hA^\dagger \hA \cong \Phi_1(\hat N_1),~~
 \hA \hA^\dagger \cong \Phi_1(\hat N_1+1) ,
\end{eqnarray}
with the structure function~(\ref{1st}). Note also that we have the
desired commutation relation
\begin{eqnarray} \label{Na1}
~[\hA,\hat N_1]\cong \hA ,
 ~~[\hA^\dagger,\hat N_1]\cong
    -\hA^\dagger .
\end{eqnarray}
From $\Phi_1(0)=0$, the number operator annihilates the
ground states:
\begin{eqnarray} \label{N0:1}
\hat N_1|0_{E/O}\rangle_1 \equiv |0_{E/O}\rangle_1
\Phi_1^{-1}(0) = 0 .
\end{eqnarray}
The $n^{th}$ excited states are generated by the formula:
\begin{eqnarray} \label{n:da}
|n_{E/O}\rangle_1 = \frac{(\hA^\dagger)^n}{\sqrt{[n]_1!}}
|0_{E/O}
    \rangle_1 ,
\end{eqnarray}
where
\begin{eqnarray} \label{[n1]}
[n]_1!\equiv \prod^n_{k=1}[k]_1= \prod_{k=1}^n \Phi_1(k) .
\end{eqnarray}
Due to Eqs.~(\ref{ada1}) and (\ref{Na1}), we have
\begin{eqnarray} \label{eigenstates}
\hat N_1|n_{E/O}\rangle_1 \cong n|n_{E/O}\rangle_1, ~~
_1\langle n_{E/O} |m_{E/O}\rangle_1 \cong \delta_{nm} .
\end{eqnarray}
Eqs.~(\ref{ada1}), (\ref{Na1}), and (\ref{N0:1}) constitute conditions
for a generalized deformed algebra which possesses a Fock space of
eigenvectors $|0_{E/O}\rangle_1,~|1_{E/O}\rangle_1,~\cdots,
|n_{E/O}\rangle_1, \cdots$ of the number operator $\hat N_1$.

We now explicitly calculate the $1^{st}$ order ground state
$|0_{E/O}\rangle_1$ of Eqs.~(\ref{N0:1}) and (\ref{n:da}),
defined by
\begin{eqnarray} \label{0:1st}
\hat A_1|0_{E/O}\rangle_1 \cong 0 \,.
\end{eqnarray}
The state $|0_{E/O}\rangle_1 $ can be written as a linear
combination of the $0^{th}$ order states,
\begin{eqnarray} \label{0:1}
|0_{E/O}\rangle_1=M_1\left( |0_{E/O}\rangle_0 + \delta
\sum_{n=1}g_n |n_{E/O}\rangle_0 \right),
\end{eqnarray}
where $M_1$ is the normalization constant of the state
$|0_{E/O}\rangle_1$ and $g_n$ are constants to be
determined by the definition of the ground
state~(\ref{0:1st}):
\begin{eqnarray} \label{0:1eq}
 \hat a_1|0_{E/O}\rangle_1\cong 0&\Longrightarrow & \delta
    \sum_n g_n\hat a_0|n_{E/O}\rangle_0+\hat a_1|0_{E/O}
        \rangle_0 \\
 && =\delta\sum_{n=0}g_{n+1} |n_{E/O}\rangle_0
        \sqrt{[n+1]}
        + \hat a_1|0_{E/O}\rangle_0 \cong 0 \nn
    ,
\end{eqnarray}
where $\hat a_0$ is given in Eq.~(\ref{a0:def}) and the
$O(\delta^0)$ equation
\begin{eqnarray} \label{an}
\hat a_0|n\rangle_0 \simeq |n-1\rangle_0\sqrt{[n]},
\end{eqnarray}
is used. The explicit form of $\hat a_1|0_{E/O}\rangle_0 $
is given by
\begin{eqnarray} \label{a1:0}
\hat a_1 |0\rangle_0 =\frac{\delta}{2\alpha}\left[-i
    \left(h(0)-\frac{\varepsilon n_0}{\alpha^2}\right)
        |0\rangle_0
    -\frac{2\epsilon h_2}{\alpha^3}\frac{\alpha+n_0}{
        b e^{-i\alpha \Omega t}}
        |1\rangle_0
 + \frac{i\sqrt{[2]_0!}\varepsilon^2}{
        4\alpha^4 b^2 e^{-2i\alpha \Omega t}}
        |2\rangle_0\right] +O(\delta^2) ,
\end{eqnarray}
where we have used
\begin{eqnarray*} \label{a:a'}
\ca \cong\hat a_0+
    \frac{i\varepsilon \delta}{2\alpha^3}\left[ \cH+n_0 +
        \frac{\varepsilon}{4\alpha^2}\left(
            \hat a_0^2+\hat a_0^{\dagger 2}\right)
       \right]
,
\end{eqnarray*}
which is derived from Eqs.~(\ref{a0:def}), (\ref{ab:sol0}),
and (\ref{T:a}).
We have also used the fact that the zeroth order excited states are
created by $\hat a_0^\dagger$.
The $0^{th}$ order algebra~(\ref{ada}) still holds for $\hat a_0$
because $\ca$ differs from $\hat a_0$ only in $O(\delta^1)$.
From Eqs.~(\ref{0:1eq}) and (\ref{a1:0}), the coefficients $g_i$ for the
$1^{st}$ order ground state~(\ref{0:1}) become
\begin{eqnarray} \label{a1:02}
 g_1 &=& \frac{i}{2\alpha}\left(h(0)-\frac{\varepsilon n_0}{
    \alpha^2}\right) ,~~~~
 g_2= \frac{\epsilon h_2(\alpha+n_0)}{\alpha^4 \sqrt{[2]_0}
    be^{-i\alpha\Omega t} }
    ,~~~~
g_3 = -\frac{i \varepsilon^2}{8\alpha^5 b^2e^{-2i\alpha
    \Omega t }}\sqrt{\frac{[2]_0}{[3]_0}},~~~~g_{n>3}=0 .
\end{eqnarray}

\subsection{General iterative formula for higher order
invariant operators}

We now develop an iterative formula which gives the $n^{th}$ order
creation and annihilation operators from the lower order operators.
The procedure in obtaining a general formula is parallel to the process
of finding the $1^{st}$ order creation and annihilation operators from
the $0^{th}$ order ones.
We try to find the series solution with the following ansatz:
\begin{eqnarray} \label{gen}
\hat A &=&e^{i \hat \Psi(t)- i \alpha \Omega t}
    \sum_{n=0}^\infty \delta^n
    \hat A_{(n)},
\end{eqnarray}
where $\hat A_{(0)}=\cA$, and the phase operator, $\hat
\Psi(t)$, and the $n^{th}$ order correction term, $\hat
A_{(n)}$, are
\begin{eqnarray}\label{corr}
\hat \Psi(t) &=& \Psi(0) +\alpha \Omega t +\Omega
\sum_{n=1}^\infty \delta^n \int^t dt'
    {\cal H}_n(t');~~~~ {\cal H}_n= \sum_{k=0}^n
        h_{nk}(t) {\cH}^k \,, \\
\hat A_{(n)} &=& \sum_{p,q=0}^\infty \left[b_{(n)}^{p,q}(t)
(\cA^\dagger)^p {\cH}^q
+c_{(n)}^{p,q}(t){\cH}^p\cA^q\right] ;\quad\quad
c_{(n)}^{p,0}=0=c_{(n)}^{p,1} ,\nn
\end{eqnarray}
with $\Psi(0)$ being an arbitrary real constant number, and
$h_{nk}(t)$, $b_{(n)}^{p,q}(t)$, and $c_{(n)}^{p,q}(t)$ are
time dependent parameters.
The operators are ordered as in the normal ordering.
A nontrivial function of operator $\cH$ is included in the phase
operator $\hat \Psi(t)$ as in the first order case.
The operator $\cH_n$ in the phase operator generates terms of the form,
$\cA$ and $\cH^m \cA$.
Therefore, such correction terms are omitted in $\hat A_{(n)}$.

As we notice in Eq.~(\ref{..LvN3}), the total derivative
$\displaystyle \frac{d \hat A_{(n)}}{dt}=
\frac{\partial\hat A_{(n)}}{\partial t}+ \frac{[\hat
A_{(n)},H]}{i\hbar}$ applies to the coefficients only:
\begin{eqnarray} \label{eq}
\frac{2i}{\Omega} \frac{d \hat A_{(n)}}{d t} \simeq
\frac{2i}{\Omega}\sum_{p,q}\left[ \dot b_{(n)}^{p,q} (\hat
    {\cal A}^\dagger)^p {\cH}^q
    +\dot c_{(n)}^{p,q}{\cH}^p
    (\cA)^q\right] .
\end{eqnarray}
To determine the parameters $h_{nk}(t)$,
$b_{(n)}^{p,q}(t)$, and $c_{(n)}^{p,q}(t)$, we need to
write $\dsp \frac{d \hat A_{(n)}}{d t}$ in terms of $\hat
A_{(k)}$ with $k<n$.
    This can be accomplished by truncating the LvN
equation~(\ref{LvN}):
\begin{eqnarray} \label{LvN:gensol}
0\equiv \frac{2i}{\Omega} \frac{d \hat A}{dt}=
\sum_{n=0}^\infty \delta^n\left[-2 \sum_{k=1}^n {\cal H}_k
    \hat A_{(n-k)}+\frac{2i}{\Omega} \frac{d \hat
    A_{(n)}}{dt}\right] .
\end{eqnarray}
If we truncate this equation at  $O(\delta^n)$, we get an iterative
equation for obtaining $\hat A_{(n)}$ from $\hat A_{(k<n)}$:
\begin{eqnarray} \label{2ndeq}
\frac{2i}{\Omega} \frac{d \hat A_{(n)}}{d
t}=\frac{2i}{\Omega}\frac{\partial}{\partial t}
 \hat A_{(n)}+[\hat A_{(n)},
    \cH]\simeq 2({\cal H} A)_{(n-1)}
    -\hat D_{(n-1)},
\end{eqnarray}
where
\begin{eqnarray*}
({\cal H} A)_{(n-1)}&=& \sum_{k=1}^n {\cal H}_k \hat
A_{(n-k)}, \\
\hat D_{(n-1)}&=&
\frac{1}{\delta}\left[\frac{2i}{\Omega}\frac{d\hat
    A_{(n-1)}}{dt}-2({\cal H}A)_{(n-2)}+ \hat D_{(n-2)}\right];
\quad\quad D_{(0)} =\frac{2i}{\delta \Omega} \frac{d \hat
A_{(0)}}{dt} .%
\end{eqnarray*}
Note that the right-hand side of Eq.~(\ref{2ndeq}) contains operators
$\hat A_{(k)}$ with $k<n$.
Therefore, the only remaining calculation to get $\dot b_{(n)}$ and
$\dot c_{(n)}$ is a simple comparison of the coefficients of the
right-hand sides of Eqs.~(\ref{eq}) and (\ref{2ndeq}).
The comparison of the coefficients of $\cH^m \ca$ determines
$h_{nk}(t)$.
After inserting this explicit results to Eq.~(\ref{2ndeq}), we can
integrate Eq.~(\ref{2ndeq}) over time to get $\hat A_{(n)}$:
\begin{eqnarray} \label{id}
\hat A_{(n)} =\frac{\Omega}{2i}\int^t dt' \left[
    2({\cal H} A)_{(n-1)}
    -\hat D_{(n-1)}\right]\, ,
\end{eqnarray}
where it is noted that only the coefficients of the
invariant operators are integrated. The calculation of
$\hat A_{(1)}$ from $\cA$ and $\cH$ in the previous
subsection is the simplest application of this formula.

\section{Summary and discussions}
We have developed a new variational perturbation method
which can be used for both the anharmonic and the double
well oscillator systems simultaneously.
    This method is based on the observation that both the
anharmonic and the double-well oscillator systems are
parity invariant and the energy eigenstates with definite
parity eigenvalues are approximately equidistanced for both
systems.
    Thus the method consists of finding the creation
and the annihilation operators which are even functions of
the position and momentum operators, and raise or lower the
energy eigenstates by two steps preserving the parity
eigenvalues.
    To do this we expand the creation and the annihilation
operators in an expansion parameter, $\delta$, which is
defined by the ratio of the length scales of the system and
the trial wavefunctions, and require them to satisfy the
LvN equation order by order in $\delta$.

The zeroth order solution of the LvN equation contains two variational
parameters.
By minimizing the energy expectation value with respect to the variation
of these parameters, we find the zeroth order energy eigenvalues for
both the anharmonic and the double-well oscillators as shown in the
Table 1.
The errors of the numerical results are small enough $\sim 10^{-2}\%$
for vast range of coupling strength, which is comparable to the $2^{nd}$
order perturbative results of other Gaussian based approximations.
The algebraic structure satisfied by the solutions provides a complete
Hilbert space constructed from the variationally-determined ground
state.
The higher order corrections are obtained by applying the standard
perturbation theory based on the zeroth order variational result.

The algebra of the double-well and the anharmonic oscillator systems to
the zeroth order in $\delta$ is given by the generalized deformed
oscillator with the structure function of the form,
\begin{eqnarray*}
H &\simeq& E_0 +  e_1 \hat N_0 , \\
\Phi_0(\hat N_0)&\simeq& \psi_1\hat N_0+\psi_2 \hat N_0^2,
\end{eqnarray*}
where $\hat N_0$ is the number operator to the zeroth order in $\delta$,
$E_0$ is the ground state energy, and $e_1$ and $\psi_i$ are real
constants determined by the LvN equation.
An interesting fact appears if one constructs the algebra to the first
order in $\delta$.
The Hamiltonian has the form:
\begin{eqnarray*} \label{en}
H \cong E_0' +  e_1' \hat N_1 + \delta e_2' \hat N_1^2,
\end{eqnarray*}
where $\hat N_1$ is the number operator to the first order in $\delta$,
and $E_0'$ and $e_i'$ are the ground state energy and constants to the
first order in $\delta$.
The energy gets corrections proportional to the square of the number
operator. On the other hand, the structure function to the $1^{st}$
order does not get $\hat N_1^3$ correction:
\begin{eqnarray*} \label{str}
\Phi_1(\hat N_1) &\cong & \psi_1' \hat N_1 + \psi_2' \hat
N_1^2 ,
\end{eqnarray*}
where $\psi_i'$ are the constants to the $O(\delta)$.
This structure function is of the same structure as that of the zeroth
order one.
This shows that the form of the structure function does not change even
if we consider the first order corrections to the zeroth order result
except for the $O(\delta)$ changes of the coefficients.
It is an interesting question if this property continue to higher
orders.

\vspace{.5cm}
\begin{acknowledgments}
This work was supported by the Korea Research Foundation Grant funded by
Korea Government(MOEHRD, Basic Research Promotion Fund)"
(KRF-2005-075-C00009;H.-C.K.) and in part by Korea Science and
Engineering Foundation Grant No. R01-2004-000-10526-0.
\end{acknowledgments} \vspace{2cm}

\begin{appendix}

\section{Variation of the effective potential }
In this appendix, we give the explicit variational calculation of
Eq.~(\ref{yc}).
Varying $V_{eff}(\bar y_0)$ with respect to $\bar y_0$, we get
\begin{eqnarray} \label{vary}
\left.\frac{2}{\hbar |\omega|\bar \Omega}\frac{d
V_{eff}(\bar y_0)}{d\bar y_0}\right|_{\bar y_0=\bar
y_c}=\left[\frac{\bar \Omega'}{\bar \Omega}\left(
    T+\frac{\lambda_r
    R}{2\bar \Omega^3}\right)
 +T'-\frac{\lambda_r R'}{4\bar \Omega^3}
 \right]_{\bar y_0=y_c}=0,
\end{eqnarray}
where $'$ denotes derivative with respect to $ \bar y_0$.
From Eq.~(\ref{dOm=0}) we have
\begin{eqnarray} \label{eqll}
\bar \Omega'&=&
 \frac{\sign(\omega^2) \bar \Omega  \frac{d}{d\bar y_0}
    \frac{2\bar y_0^2}{T}
    +\lambda_r \frac{d}{d\bar y_0}\frac{R}{T}}{
 3\bar\Omega^2-\sign(\omega^2)
 \left(1+\frac{2\bar y_0^2}{T(\bar y_0)}\right)}\\
&=&\frac{4\bar y_0}{T}\, \frac{\sign(\omega^2) \bar \Omega+
 \frac{
    2\sign(\omega^2)\bar y_0^2\bar \Omega+\lambda_r R}{T} C(1\mp
    \bar y_0^2 e^{\bar y_0^2}C)+
    \frac{2\lambda_r}{1\pm e^{-\bar y_0^2}}
    \left(3+2\bar y_0^2+\bar y_0^2(3+\bar y_0^2) C
    \right)}{3\bar \Omega^2-\sign(\omega^2)
    \left(1+\frac{2\bar y_0^2}{T(\bar
y_0)}\right)}    \nn  ,
\end{eqnarray}
where we have used
\begin{eqnarray*}
T' &=&-4 \bar y_0 C\left(1\mp \bar y_0^2 e^{\bar y_0^2}
C\right)= \mp \frac{4\bar y_0}{e^{\bar y_0^2}\pm 1}\left(1-
    \frac{\bar y_0^2}{1\pm e^{-\bar
    y^2}}\right), \nn \\
R' &=&\frac{8\bar y_0}{1\pm e^{-\bar y_0^2}}\left(3+2\bar
    y_0^2+\bar y_0^2(3+\bar y_0^2) C\right)  .
 \nn
\end{eqnarray*}
Thus, the variational equation~(\ref{vary}) becomes an equation for
$y_c$:
\begin{eqnarray} \label{y:1}
&&4\bar y_0\left[\pm \frac{1+\frac{\lambda_r R}{2\bar
    \Omega^3 T}}{3\bar \Omega^2-\sign(\omega^2) \left(1+ \frac{2 \bar
    y_0^2}{T}\right)}
-\left(1-\frac{( 2\sign(\omega^2)\bar \Omega \bar y_0^2
+\lambda_r
    R)\left(1+\frac{\lambda_r R}{2\bar \Omega^3 T}\right)}{
    3\bar \Omega^3-\sign(\omega^2) \left(1+ \frac{2 \bar
    y_0^2}{T}\right)\bar \Omega}\right)
    C\left(1- \frac{\bar y_0^2}{
    1\pm e^{-\bar y_0^2}}\right) \right. \\
&&\left. +2\lambda_r\left(\frac{1+\frac{\lambda_r R}{2\bar
    \Omega^3 T}}{3\bar \Omega^3-\sign(\omega^2) \left(1+ \frac{2 \bar
    y_0^2}{T}\right)\bar \Omega}-\frac{1}{4\bar
    \Omega^3}\right) \frac{3+2\bar y_0^2 + \bar y_0^2(3+
    \bar y_0^2)C}{1\pm e^{-\bar y_0^2}}
 \right]_{\bar y_0=y_c} =0 .\nn
\end{eqnarray}
By numerical calculation, it can be shown that this equation allows a
unique positive solution for $y_c$ for every positive $\lambda_r$.

\section{Structure of the Hamiltonian in $\delta-$expansion}
In this appendix, we study the explicit behaviors of the coefficients
$h_i$ in $\delta-$expansion of the Hamiltonian~(\ref{H1:2}), which can
be rewritten in the form:
\begin{eqnarray} \label{H:3}
H &=& \frac{\hbar \Omega}{2}\left(
    \frac{1}{2}T_2^0+\frac{\epsilon h_1 }{\epsilon_0^2} T_0^2
    +\frac{\varepsilon h_2}{2}
    T_0^4\right)+V(x_0)\equiv \frac{\hbar \Omega}{2}\cH +E_0;
    \\
h_1&=&
    \frac{\epsilon_0^2}{\epsilon^2}\left(\frac{\omega^2}{\Omega^2}+\frac{\lambda
    x_0^2}{\Omega^2}\right)=\frac{\epsilon_0^2\nu^4}{
    \epsilon^2
    \bar \Omega^2}\left(\sign(\omega^2)+\frac{2\lambda_r
    \bar y_c^2}{\bar \Omega}\right), \quad\quad
h_2=\frac{\lambda
  \hbar}{\Omega^3\varepsilon\epsilon^2}=\frac{2\nu^6}{\varepsilon
  \epsilon^2}\frac{\lambda_r}{\bar \Omega^3}
  ,\nn
\end{eqnarray}
where $E_0$ is the ground state energy and $\Omega$, $\epsilon$,
$\varepsilon$, and $x_0$ are given by the values determined by the
zeroth order variational approximation in section III.

The $\lambda_r \rightarrow 0$ limit corresponds to the $\epsilon
\rightarrow 0$ limit or $\varepsilon \rightarrow 0$ limit depending on
the sign of $\omega^2$.
Therefore, the analysis of the small $\lambda_r$ limit of the
variationally determined Hamiltonian, which was done in
subSec.~(\ref{sub1}), provides the order of magnitude of the
coefficients $h_i$ with respect to $\epsilon$ and $\varepsilon$.
Explicitly, the numbers $h_{1}$ and $h_2$ are of $O(\delta^0)$.
For example, in the $\lambda_r \rightarrow 0$ limit of the zeroth order
approximation for ground state we have,
\begin{eqnarray*}
h_1=\frac{\epsilon_0^2}{\epsilon^2}\left(\frac{\omega^2}{\Omega^2}+\frac{\lambda
    x_0^2}{\Omega^2}\right)
    &=&\left\{\begin{tabular}{ll}
    $\dsp 1-\frac{\varepsilon}{2}-\frac{3\varepsilon^2}{4}
        +O(\varepsilon^3)$,&
         ~~$\omega^2>0,~~~\lambda_r \ll 1,$ \vspace{.1cm}\\
    $\dsp-\frac{3\sqrt{2}}{\Omega_r^3}
        +O(\epsilon^2)$, &
    $\omega^2<0,~~~\lambda_r \ll 1,$ \\
    \end{tabular}\right\}\Longrightarrow O(\delta^0) , \\
h_2=\frac{\lambda \hbar
    }{\Omega^3\epsilon^2\varepsilon}
    &=&\left\{\begin{tabular}{cc}
    $\dsp-1+O(\varepsilon)$,&
         $\omega^2>0,~~~\lambda_r \ll 1,$ \vspace{.1cm}\\
    $\dsp 1+\frac{11}{2}\epsilon^2 +O(\epsilon^3)
        $, &
    ~~$\omega^2<0,~~~\lambda_r \ll 1,$ \\
    \end{tabular}\right\}\Longrightarrow O(\delta^0) .
\end{eqnarray*}
Note also that the differences $\epsilon (h_1-1)$ and
$\varepsilon (h_2-1)$ are of $O(\delta)$:
\begin{eqnarray} \label{coef3}
\frac{\epsilon(h_1-1)}{\epsilon_0^2}=
\frac{1}{\epsilon}\left(\frac{\omega^2}{\Omega^2}
    +\frac{\lambda
    x_0^2}{\Omega^2}-\frac{\epsilon^2}{\epsilon_0^2}\right)
    &=&\left\{\begin{tabular}{cc}
    $-\varepsilon^2+O(\varepsilon^3)$,&
         $\omega^2>0,~~~\lambda_r \ll 1,$ \vspace{.1cm}\\
    $\dsp-\frac{6\sqrt{2}-1}{\Omega_r^3} \epsilon
        +O(\epsilon^3)$, &
    $\omega^2<0,~~~\lambda_r \ll 1,$ \\
    \end{tabular}\right\}\Longrightarrow O(\delta) , \\
\frac{\varepsilon(h_2-1)}{2}=\left(\frac{\lambda \hbar
    }{2\Omega^3\epsilon^2}-\frac{\varepsilon}{2}\right)
    &=&\left\{\begin{tabular}{cc}
    $-\varepsilon+O(\varepsilon^2)$,&
         $\omega^2>0,~~~\lambda_r \ll 1,$ \vspace{.1cm}\\
    $\dsp \frac{11}{4}\epsilon^2 +O(\epsilon^3)
        $, &
    $\omega^2<0,~~~\lambda_r \ll 1,$ \\
    \end{tabular}\right\}\Longrightarrow O(\delta) .\nn
\end{eqnarray}
If we define $\bar h_1$ and $\bar h_2$ as
\begin{eqnarray} \label{barh}
\bar h_1&=& \frac{\epsilon(h_1-1)}{\delta\epsilon_0^2},
 \quad \quad \bar h_2=\frac{\varepsilon(h_2-1)}{2\delta}
  ,
\end{eqnarray}
it is clear that $\epsilon \bar h_1$ and $\varepsilon \bar h_2$ are of
$O(\delta)$.

\section{Operator basis for perturbative expansion}
In this appendix, we introduce an operator basis for calculational
convenience.
Our basic elements $T_m^n$ are defined as the normalized sum of all
possible terms containing $m$ factors of $\hat \pi$ and $n$ factors of
position operators (with the combination of $\hat y$ and $(\hat
y^2-y_0^2$)).
$T_m^n$ is thus a totally symmetric Hermitian operator of
$O(\epsilon^0)$ and $O(\varepsilon^0)$,
\begin{eqnarray} \label{T}
T_{m}^{2n} &=&\frac{\epsilon^{n}
  }{2^{m/2}} \sum_{j=0}^m
  \left(\begin{tabular}{c} $m$\\$j$
  \end{tabular}\right)\hat \pi^j (y^2-y_0^2)^n
    \hat \pi^{m-j}, \\
T_{m}^{2n+1}&=& \frac{\epsilon^{n+1}
  }{2^{(m-1)/2}} \sum_{j=0}^m
  \left(\begin{tabular}{c} $m$\\$j$
  \end{tabular}\right)\hat \pi^j y(y^2-y_0^2)^n
    \hat\pi^{m-j} . \nonumber
\end{eqnarray}
For two reasons, this seems to be a natural basis with which to express
operators for bi-centered systems.
First, $T_m^n$ contains positive powers of $\pi$ and $y$, so it is
useful for constructing generalized expansions of operators as we do in
Eq.~(\ref{f:approx}) in function space.
Note that we can expand every nonsingular operators in terms of $T_m^n$,
regardless of whether they are symmetric or not. For example,
\begin{eqnarray*}
\epsilon^3 \pi^2 y^3= \frac{1}{2\sqrt{2}}\left(\epsilon
T_2^3
    +\frac{\varepsilon}{2} T_2^1-3\epsilon^2 T_0^1\right)
    -i\left(\frac{3\epsilon^2}{2}T_1^2+\sqrt{2}
    \delta T_1^0\right) ,
\end{eqnarray*}
where $\delta$ appears as a natural expansion parameter. Secondly,
$T_{m}^n$ satisfies useful set of commutation and anticommutation
relations. Commutator of $T_m^n$ with $\hat \pi$ and $\hat y$ has the
effect of lowering the orders of the operator and displacing by
$y_0=\sqrt{\varepsilon/(2\epsilon^2)}$:
\begin{eqnarray} \label{com:0}
~[T_m^{n}, \hat y]&=& - \sqrt{2} m i T_{m-1}^{n} , \\
~[T_m^{2n}, \hat \pi]&=& \sqrt{2} n i T_{m}^{2n-1}, ~~
 [T_m^{2n+1}, \hat \pi]= \sqrt{2}i (2n+1)\epsilon T_m^{2n}+
 \sqrt{2}i n \varepsilon
    T_{m}^{2(n-1)}  , \nonumber
\end{eqnarray}
where the last term in the second line containing $\varepsilon$
represents the effect of bi-centered property of the system.
Anticommutator of $T_m^n$ with $\hat \pi$ or $\hat y$ has an effect of
raising the operator power and displacing by $y_0$:
\begin{eqnarray} \label{rules}
 \{T_m^{2n}, \epsilon \hat y\}_+ &=&\sqrt{2} T_m^{2n+1},~~
    \{T_m^{2n+1}, \epsilon \hat y\}_+ =\sqrt{2}
    \left(2\epsilon T_m^{2n+2}+\varepsilon
     T_m^{2n}\right), \\
 \{T_m^{n}, \hat \pi\}_+&=& \sqrt{2} T_{m+1}^{n}. \nonumber
\end{eqnarray}
When one calculates the commutator and anticommutator between higher
polynomials, the following identities are helpful:
\begin{eqnarray} \label{comm}
[A,BC+CB]&=&\{[A,C],B\}_+ +\{[A,B],C\}_+,~~\{A,BC+CB\}_+
=\{\{A,B\}_+,C\}_+ + [[A,C],B]  \\
 ~\{A, BC\}&=& [A,B]C+ B\{A,C\}= \{A, B\}C- B[A,C] \nn .
\end{eqnarray}
Similar operators as $T_m^n$ with $y_0=0$ for anharmonic oscillator were
used in Ref.~\cite{bender,bender2} and these operators have played a
central role in studying the finite-element lattice approximation, the
operator ordering, the Hahn's polynomial, and the analysis of exact
solutions of operator differential equation.

We list some of higher order formulae for the commutators with quadratic
and quartic operators,
\begin{eqnarray*}
~[T_m^{2n},T_2^0] &=&
4niT_{m+1}^{2n-1},~~[T_m^{2n+1},T_2^0]=
    4i(2n+1)\epsilon T_{m+1}^{2n}+4i\varepsilon nT_{m+1}^{2(n-1)}
    ,\\
~[T_m^{2n},T_0^2]&=&-2miT_{m-1}^{2n+1},~~[T_m^{2n+1},T_0^2]=
    -2mi(2\epsilon T_{m-1}^{2n+2}+\varepsilon T_{m-1}^{2n}), \\
~[T_m^{n},T_1^1]&=&2i(n-m)\epsilon
T_m^{n}+2i[n/2]\varepsilon
T_m^{n-2},~~ \\
~[T_m^{2n},T_0^4]&=&-4miT_{m-1}^{2n+3}+2i\epsilon(m)_3T_{m-3}^{2n+1},\\
~[T_m^{2n+1},T_0^4]&=&-4mi(2\epsilon
T_{m-1}^{2n+4}+\varepsilon
T_{m-1}^{2n+2})+2i(m)_3(2\epsilon^2 T_{m-3}^{2n+2}+\delta
T_{m-3}^{2n}) ,
\end{eqnarray*}
and formulae for the anticommutators with quadratic and quartic
operators,
\begin{eqnarray*}
~\{T_m^{n},T_2^0\}_+&=& 2T_{m+2}^{n}-(n)_2\epsilon
    T_m^{n-2}-2([n/2])_2\varepsilon T_m^{n-4} , \\
\{T_m^{n}, T_0^2\}_+&=& 2T_m^{n+2}-m(m-1)\epsilon T_{m-2}^n, \\
\{T_m^{2n},T_1^1\}_+&=& 2T_{m+1}^{2n+1}+2mn\epsilon
    T_{m-1}^{2n-1},\\
\{T_m^{2n+1},T_1^1\}_+&=&4\epsilon
T_{m+1}^{2n+2}+2\varepsilon
    T_{m+1}^{2n}+2m(2n+1)\epsilon^2 T_{m-1}^{2n}+2mn\delta
    T_{m-1}^{2(n-1)}, \\
\{T_m^{n},T_0^4\}_+&=& 2T_{m}^{n+4}-2(m)_2(3\epsilon
T_{m-2}^{n+2}
    +\varepsilon T_{m-2}^{n})+\frac{\epsilon^2}{2}(m)_4
    T_{m-4}^{n} ,
\end{eqnarray*}
where $(m)_n=m(m-1)(m-2)\cdots (m-n+1)$. All of these formulae are
obtained by using (\ref{com:0}) and (\ref{rules}), and (\ref{comm}).

\section{Some formulae for higher order calculation}
In this appendix, we list some formulae which are needed in evaluating
higher order invariant operators.

The operator $\hat B$ of Eq.~(\ref{B}) is an $O(\delta^0)$ operator
defined by
\begin{eqnarray*}
\hat B = 2\epsilon
(\cH+n_0)-\frac{\alpha}{2i}(\ca-\ca^\dagger) .
\end{eqnarray*}

With this and Eq.~(\ref{comm}), we have,
\begin{eqnarray} \label{A0:Hm}
~[\ca, \hat B]&=& 2\epsilon[\ca, \cH]+\frac{\alpha}{2i}
    [\ca, \ca^\dagger],\\
\{ \ca, \hat B\}_+ &=& 4\epsilon(\cH +n_0)\ca+ 2\epsilon
[\ca,
    \cH]+i\alpha
    \ca^2-\frac{i\alpha}{2}\{\ca,\ca^\dagger\}_+ ,\nn
\end{eqnarray}
where
\begin{eqnarray}
 ~[\ca, \cH]&= & 2 \alpha \ca -
    \delta\left[\frac{2ih_1}{\epsilon_0^2}
    +2 h_2\left(2i T_0^4
    -\frac{\{T_0^2,T_1^1\}}{\alpha}\right)\right]
 +\delta^2 \left[\frac{4\epsilon \bar h_1}{\alpha\delta}
    T_1^1 -\frac{4i\left(\epsilon\bar h_1
        +\varepsilon \bar h_2\right)}{\delta} T_0^2
   \right]
   \, , \\
\frac{1}{4}[\ca, \ca^\dagger]&= &  \nu^2 \varepsilon
     +\frac{4 \epsilon^2 (\cH +n_0)}{\alpha}
     +\frac{2\delta}{\alpha} \left[
     (\varepsilon-2\epsilon^2\bar h_1) T_0^2
     -\epsilon h_2 (T_0^2)^2\right]
 ,\nn\\
\frac{1}{4}\{\ca,\ca^\dagger\}_+&=&\frac{3\epsilon^2}{2}+
 \varepsilon(\cH+n_0)+\frac{2\epsilon^2}{\alpha^2}
        (\cH+n_0)^2
-\frac{\delta h_1}{\epsilon_0^2}T_0^2
 -\frac{\delta\epsilon h_2}{\alpha^2}\{\cH+n_0,
    (T_0^2)^2\}_+
 \nn\\
&+&\delta^2\left[-\frac{\varepsilon \bar
    h_2}{\delta}+ \frac{2(\epsilon \bar h_1)^2}{\alpha^2}
    \right]T_0^4 -\frac{2\delta (\epsilon^2\bar
    h_1)}{\alpha^2}\{\cH+n_0,T_0^2\}_+
 +\frac{\delta^2 h_2^2}{2\alpha^2}T_0^8
 + \frac{2\delta^2h_2(\epsilon \bar h_1)}{\alpha^2} T_0^6
.\nn
\end{eqnarray}
where we have used $T_1^3= \{T_0^2,T_1^1\}/2$ to write $T_1^3$ in terms
of $\cA$, $\cA^\dagger$, and $\cH$.
%
The square of $\hat B$ can be written as
\begin{eqnarray} \label{b2}
\hat B^2= 4\epsilon^2(\cH+n_0)^2+
 i\epsilon\alpha \{\cH +n_0, \ca
 -\ca^\dagger\}_+ +\frac{\alpha^2 }{4}\left(
    \{\ca, \ca^\dagger\}_+
    -\ca^2-\ca^{\dagger 2}\right) .
\end{eqnarray}

We now list formulae needed to compute the commutators and
anticommutators for the higher order corrections:
\begin{eqnarray} \label{A0:Hm}
~[\ca, \hat B]&\simeq & 4\alpha\epsilon\ca
    -2i\alpha \left(\nu^2 \varepsilon
    +\frac{4\epsilon^2 (\cH+n_0)}{\alpha}\right),\\
\{\ca+\ca^\dagger, \hat B\}_+ &\simeq & 4\epsilon\left[
    (\cH+n_0+\alpha)\ca + \ca^\dagger(\cH +n_0+\alpha)\right]
    + i\alpha(
    \ca^2-\ca^{\dagger 2}), \nn \\
\alpha^4T_0^4&\simeq& \hat B^2 \simeq
    \frac{3\epsilon^2\alpha^2}{2}+\varepsilon
    \alpha^2 (\cH+n_0)
    +6\epsilon^2(\cH+n_0)^2-\frac{\alpha^2(\ca^2
    +\ca^{\dagger 2})}{4} \nn\\
    && + 2i\epsilon \alpha\left[(\cH+n_0+\alpha)\ca
        -\ca^\dagger(
    \cH+n_0+\alpha)\right] , \nn \\
\ca \cH &\simeq & (\cH+2\alpha) \ca, \quad \quad
\cH \ca^\dagger \simeq \ca^\dagger (\cH + 2\alpha), \nn\\
 ~[\cH+n_0,T_0^4] &=&-4i T_1^3 = -2i \{T_0^2, T_1^1\}_+
    \simeq -\frac{i}{\bar \alpha^2}\{\hat B,
        \ca+\ca^\dagger\}_+\nn \\
&\simeq &\frac{1}{\alpha} (\ca ^2-\ca^{\dagger 2})-\frac{4i
    \epsilon}{\alpha^2} \left[
    (\cH +n_0+\alpha) \ca +\ca^\dagger
    (\cH+n_0+\alpha)\right] , \nn\\
~\{\cH+n_0,T_0^4\}_+ &\simeq & \frac{1}{\alpha^2}\left[
    2\varepsilon (\cH+n_0)^2+ 3\epsilon^2 (\cH+n_0)\left(
        1+\frac{4(\cH+n_0)^2}{\alpha^2}\right)\right] \nn\\
&-&\frac{1}{2\alpha^2}\left[(\cH+n_0+2\alpha) \ca^2+
    \ca^{\dagger 2}(\cH+n_0+2\alpha)\right]
+ \frac{4i\epsilon}{\alpha^3} \left[ (\cH+n_0+\alpha)^2
    \ca -\ca^\dagger (\cH+n_0+\alpha)^2\right]
. \nn
\end{eqnarray}

To find the structure function to $O(\delta)$, we need the commutator
and the anticommutator in terms of invariant operators to the same
order:
\begin{eqnarray} \label{..}
\frac{1}{4}[\ca, \ca^\dagger]&\cong& \nu^2 \varepsilon
     +\frac{4\epsilon^2(\cH+n_0)}{\alpha}
      -\frac{3\epsilon^3 \delta h_2}{
        \alpha^3}\left(1+\frac{4(\cH+n_0)^2}{\alpha^2}\right)
    \\
&-&\frac{2\delta h_2}{\alpha^3} \left[
     \frac{i\varepsilon\alpha}{2}(
            \ca-\ca^\dagger)
   -\frac{\epsilon}{4}\left(\ca^2+\ca^{\dagger 2}\right)
   +\frac{2i\epsilon^2}{\alpha}\left(
   (\cH+n_0+\alpha)\ca-\ca^\dagger
   (\cH+n_0+\alpha)\right)
    \right]
 ,\nn\\
\frac{1}{4}\{\ca,\ca^\dagger\}_+
 &\cong&\frac{3\epsilon^2}{2}+
    \varepsilon(\cH+n_0)+\frac{2\epsilon^2}{\alpha^2}
        (\cH+n_0)^2
-\frac{2\epsilon \delta h_1}{\epsilon_0^2\alpha^2}(\cH+n_0)
    \nn \\
&-&\frac{3\delta \epsilon^3h_2}{\alpha^4}(\cH+n_0)
    \left(1+\frac{4(\cH+n_0)^2}{\alpha^2}\right)
+\frac{\delta \epsilon h_2}{2\alpha^4}\left[
    (\cH+n_0+2\alpha) \ca^2+ \ca^{\dagger 2} (\cH+n_0+2
    \alpha)\right] \nn \\
&-&\frac{4i\delta \epsilon^2h_2}{\alpha^5}\left[
    (\cH+n_0+\alpha)^2 \ca- \ca^{\dagger} (\cH+n_0+
    \alpha)^2\right]
    -\frac{i\delta
    h_1}{2\epsilon_0^2\alpha}(\ca-\ca^\dagger) , \nn \\
~[\ca,\cH]&\cong & 2\alpha \ca
    -\frac{2ih_1\delta}{\epsilon_0^2}
 -\frac{4ih_2\delta}{\alpha^4}\left\{
    \frac{3\epsilon^2\alpha^2}{2}+\varepsilon \alpha^2(\cH
    +n_0) \right. \nn \\
  &+&\left.6\epsilon^2(\cH+n_0)^2-\frac{\alpha^2}{2}\ca^2+
  i\epsilon \alpha\left[3(\cH+n_0+\alpha)\ca-\ca^\dagger(
    \cH+n_0+\alpha)\right] \right\}
.\nn
\end{eqnarray}

The commutators between the zeroth order invariant operator and the
first order correction term are
\begin{eqnarray} \label{comm3}
~[\ca, \hat a_{(1)}^\dagger] &\simeq&
 \frac{4i h_2}{\alpha^3}\left\{
    \varepsilon(\alpha\ca+2 \nu^2\ca^\dagger)
 +\frac{i\epsilon}{2}\ca^2+\frac{4\epsilon^2}{\alpha}
    \left[3(\cH+n_0+ \alpha) \ca+2\ca^\dagger
    (\cH+n_0+\alpha)\right]\right\},
     \\
~[\hat a_{(1)},\ca^\dagger] &\simeq &-
  \frac{4i h_2}{\alpha^3}\left\{
    \varepsilon(\alpha\ca^\dagger+2 \nu^2\ca)
 -\frac{i\epsilon}{2}\ca^2
 +\frac{4\epsilon^2}{\alpha}
    \left[2(\cH+n_0+ \alpha) \ca+3\ca^\dagger
    (\cH+n_0+\alpha)\right]\right\}.
    \nn
\end{eqnarray}
Summing the two, we have
\begin{eqnarray} \label{commSum}
&&\frac{[\ca, \hat a_{(1)}^\dagger]+[\hat
a_{(1)},\ca^\dagger]}{4}
    \\
&&\quad \simeq \frac{i h_2}{\alpha^3}\left\{
    \frac{(\alpha^2-2\alpha \nu^2)\varepsilon}{\alpha}
    (\ca-\ca^\dagger)
 +\frac{i\epsilon}{2}(\ca^2+\ca^{\dagger 2})
    +\frac{4\epsilon^2}{\alpha}\left[(\cH+n_0+\alpha) \ca
        -\ca^\dagger (\cH+n_0+\alpha)\right] \right\} .
    \nn
\end{eqnarray}

The anticommutators between the zeroth order invariant operator and the
first order correction term are
\begin{eqnarray} \label{comm3}
~\{\ca, \hat a_{(1)}^\dagger\}_+ &\simeq&
 \frac{i}{\alpha}\left[h(\cH)+\frac{4 \varepsilon h_2}{\alpha}
 +\frac{48\epsilon^2h_2}{\alpha^3}(\cH+n_0+\alpha)
 \right]\ca \\
&-& \frac{2\epsilon
 h_2}{\alpha^4} (\cH+n_0+2\alpha)\ca^2 +
 \frac{4i h_2}{\alpha^3}\ca^\dagger\left[
   \nu^2 \varepsilon+\frac{19\epsilon^2}{2}+
   \left(\frac{4\epsilon^2}{\alpha}+\varepsilon\right)
    (\cH+n_0)+\frac{2\epsilon^2}{\alpha^2}
        (\cH+n_0)^2\right],
    \nn \\
~\{\hat a_{(1)}, \ca^\dagger\}_+ &\simeq&
 -\frac{i}{\alpha}\ca^\dagger
    \left[h(\cH)+\frac{4 \varepsilon h_2}{\alpha}
 +\frac{48\epsilon^2h_2}{\alpha^3}(\cH+n_0+\alpha)
 \right] \\
&-& \frac{2\epsilon
 h_2}{\alpha^4} \ca^{\dagger 2}(\cH+n_0+2\alpha) -
 \frac{4i h_2}{\alpha^3}\left[
   \nu^2 \varepsilon+\frac{19\epsilon^2}{2}+
   \left(\frac{4\epsilon^2}{\alpha}+\varepsilon\right)
    (\cH+n_0)+\frac{2\epsilon^2}{\alpha^2}
        (\cH+n_0)^2\right]\ca
    \nn.
    \nn
\end{eqnarray}
Summing the two, we have
\begin{eqnarray} \label{commSum}
&&\frac{\{\ca, \hat a_{(1)}^\dagger\}_++\{\hat
a_{(1)},\ca^\dagger\}_+}{4}
    \\
&&\quad\simeq
\frac{ih_1}{2\alpha\epsilon_0^2}\left(\ca-\ca^\dagger\right)
 +\frac{4i\epsilon^2 h_2}{\alpha^5}
    \left[\left(\cH+n_0+\alpha\right)^2\ca -
    \ca^\dagger\left(\cH+n_0+\alpha\right)^2\right]\nn \\
&&\quad \quad-\frac{\epsilon
    h_2}{2\alpha^4}\left[(\cH+n_0+2\alpha)\ca^2
   +\ca^{\dagger 2}(\cH+n_0+2\alpha)\right] .\nn
\end{eqnarray}

The commutator between the first order correction term to the
annihilation operator and $\cH$ becomes
\begin{eqnarray} \label{..2}
~[a_{(1)}, \cH]=\frac{2\epsilon h_2}{\alpha^3}\,\ca^\dagger
\left(\cH +n_0+\alpha\right)-\frac{4i h_2}{\alpha^2}\ca^2 .
\end{eqnarray}

\end{appendix}


\vspace{4cm}

\end{document}